\documentclass[aps,prb,amsmath, amssymb, english, twocolumn,reprint, floatfix, superscriptaddress,longbibliography]{revtex4-2}

\usepackage[mode=buildnew]{standalone}% requires -shell-escape
\usepackage[caption=false]{subfig}
\usepackage{placeins}

\usepackage{bbm}
\usepackage{comment}
\usepackage{array}                            
\usepackage{ulem}                                             
\usepackage{hhline}
\usepackage{tikz}
\usepackage{pgfplots}
\usepackage{microtype}
\usepackage{dsfont}
\usepackage{epstopdf}
\usepackage{epsfig} 
\usepackage[applemac]{inputenx}     
\usepackage{amsmath}
\usepackage{MnSymbol}
\usepackage{floatrow}
\usepackage{tabularx}
\usepackage[colorlinks=true,
            urlcolor=blue,
            citecolor=blue,
            linkcolor=blue,
            breaklinks=true]{hyperref}

\pgfplotsset{compat=1.15}
\usepackage{xurl}

\hypersetup{
    colorlinks,
    citecolor=blue,
    filecolor=blue,
    linkcolor=blue,
    urlcolor=blue
}

\newcommand{\bq}{{\mathbf{q}}}

\usepackage{mathtools}

\usepackage[dvipsnames]{xcolor}

\def\equationautorefname~#1\null{Eq.~(#1)\null}

\allowdisplaybreaks

\graphicspath{{./Images/}}

\begin{document}
%\title{Impact of nonlocal spatial correlations for different lattice geometries}
\title{Magnetic instabilities in three- and higher-dimensional Hubbard models with different lattice geometries: Impact of nonlocal spatial correlations}
\author{Marvin Leusch}
\affiliation{Institute of Theoretical Physics, University of Hamburg, 20355 Hamburg, Germany}
\author{Alessandro Toschi}
\affiliation{Institute of Solid State Physics, TU Wien, A-1040 Vienna, Austria}
\author{Andreas Hausoel}
\affiliation{Institute for Theoretical Solid State Physics, Leibniz Institute for Solid State and Materials Research Dresden, Helmholtzstrasse\ 20, 01069 Dresden, Germany}
\author{Giorgio Sangiovanni}
\affiliation{Institut f{\"u}r Theoretische Physik und Astrophysik and W{\"u}rzburg-Dresden Cluster of Excellence ctd.qmat, Universit{\"a}t W{\"u}rzburg, 97074 W{\"u}rzburg, Germany}
\author{Georg Rohringer}
\affiliation{Theory and Simulation of Condensed Matter, Department of Physics, King's College London, The Strand, London WC2R 2LS, United Kingdom}
\affiliation{Institute of Theoretical Physics, University of Hamburg, 20355 Hamburg, Germany}
\date{\today} 

\pacs{}

\begin{abstract}
We analyze the impact of the lattice geometry on the 
%\sout{phase} \at{}
thermodynamic transition to magnetically ordered phases in strongly interacting electron systems for various Bravais lattices in three, four and five dimensions, including both local and nonlocal correlation effects.
In a first step we use the dynamical mean-field theory (DMFT), which takes into account purely local correlations, to calculate the magnetic susceptibilities of the Hubbard model on three- (3d-sc), four- (4d-sc), and five-dimensional (5d-sc) simple cubic/hypercubic, as well as on three-dimensional body- (bcc) and face-centered (fcc) cubic lattices, and determine the transition temperature to the corresponding magnetically ordered state.
In a second step, we exploit the dynamical vertex approximation (D$\Gamma$A), a diagrammatic extension of DMFT, to include the effect of nonlocal correlations which are particularly important in the vicinity of the corresponding phase transition.
For the bipartite 3d-sc, 4d-sc, 5d-sc and bcc lattices nonlocal fluctuations lead to a substantial reduction of the DMFT transition temperature consistent to the overall tendency of mean-field approaches to overestimate the stability of ordered phases.
As expected, the magnitude of the difference between the DMFT, being exact in the limit of large connectivity/dimensions, and D$\Gamma$A transition temperatures decreases with increasing coordination number. On a more practical perspective, these results also provide a reasonable guidance to evaluate the expected overestimation of the DMFT ordering temperature for different material geometries.
%\sout{of the lattice as DMFT becomes numerically exact in the limit of infinite dimensions.}
%\sout{This is also reflected in the extent to which specific sum rules are violated in DMFT due to its inability to capture nonlocal correlations.}
%\gr{I would like to keep the previous sentence. @Ale: Did you remove it because the situation is not clear for the fcc lattice?}
For the fcc lattice, on the other hand, the ordered phase observed in DMFT vanishes completely within D$\Gamma$A which is consistent with the existence of strong geometric frustration in this lattice.
\end{abstract}

\maketitle

\section{Introduction}
\label{sec:Intro}

The properties of crystalline materials are are determined by two main ingredients, namely the type of periodic lattice, which is formed by the atomic ions and the interaction between the electrons which move in this positively charged background.
Even if we consider a static lattice within the well-known Born-Oppenheimer approximation\cite{Born1927}, which acts as a stage for the dynamics of the electrons, the resulting many-body problem is still very hard to tackle theoretically.
In this situation it is of uttermost importance to select the appropriate numerical (or analytical) tools for the theoretical analysis.
The actual choice is dictated by the relative importance of the two main above-mentioned ingredients for the given system, i.e., the structure of the lattice and the interaction between the electrons.
In the case of sufficiently weak interactions the physics is dominated by the structure of the underlying lattice and can be captured by low-order perturbation theory such as GW\cite{Reining2018,Held2011} or by an effective single-particle description as provided by density functional theory\cite{Hohenberg1964,Chayes1985}.
The electronic state is typically well described by Bloch wave functions and the electrons are completely delocalized over the entire lattice.
In the opposite limit, where the interaction between two electrons at one lattice site is much larger than the corresponding tunneling probability between the lattice sites, the electronic charge becomes localized in real space forming, for instance, a correlation-driven (Mott) insulating state.
Strong-coupling perturbation theory\cite{Pairault1998,Pairault2000} is one of the methods which can treat this situation quite well, being obviously not adequate for the metallic side of the system.
The intermediate case, where both the lattice structure as well as the interactions between the electrons are equally relevant, is the most challenging situation for the theoretical treatment of a many-body problem.

In this respect, the dynamical mean-field theory\cite{Metzner1989,Georges1992a,Georges1996,Held2007} (DMFT) has represented a huge step forward as it is able to capture all purely local correlations of interacting lattice electrons, allowing for a unified description of the entire coupling regime.
Among the biggest successes of DMFT are the accurate quantitative description of the correlation-driven Mott metal-to-insulator transition in model systems such as the single-band Hubbard model\cite{Hubbard1963,Kanamori_1963,Gutzwiller_1963,Georges1992} and real materials\cite{Imada1998} such as transition metal oxides\cite{McWhan1970,McWhan1973}, the uncovering of orbital selective Mott transitions\cite{Anisimov2002}, the identification of peculiar correlation-driven features in the spectral function\cite{Byczuk2007} and in the specific heat \cite{Toschi2009} or 
%\at{Ich w\"urde hier auch gewisse rezentere DMFT Papers \"uber Correlated Topology, zB von Giorgio, inkludieren}
%\gs{We could add: ``
topological transitions without gap closing \cite{amaricciPRL2015}, the recently discovered orbital-magnetic-field-driven insulator-to-metal transition in the Hubbard-Hofstadter model\cite{Rohringer2024}.
Of particular interest are also the investigation of magnetic quantum phase transitions\cite{Adler2024a} and DMFT studies of magnetically ordered phases and the associated phase transitions which are difficult to tackle theoretically even at weak to intermediate coupling\cite{DelRe2025}. 
%\sout{, where DMFT allows for a comprehensive description in the entire coupling regime.} 
%\sout{In this respect,} 
In fact, DMFT has advanced our understanding of the magnetism in iron and nickel\cite{Lichtenstein2001,Hausoel2017} or the ferro-orbital order in LaTiO$_3$\cite{Anisimov1997}.
On the model level, DMFT correctly captures the crossover from a weakly coupled Slater to a strongly coupled Heisenberg antiferromagnet in the single-band Hubbard model on a three-dimensional bipartite cubic lattice\cite{Rohringer2016,DelRe2021,Reitner2025}.

In spite of its achievements, DMFT has serious limitations where the most notable one is arguably its intrinsic inability to capture nonlocal correlation effects. 
This leads, for instance, to an overestimation of the transition temperature to magnetically ordered phases in three spatial dimensions\cite{Rohringer2011} or the prediction of  a spurious phase transition in two dimensions\cite{Schafer2015}, which is forbidden by the Mermin-Wagner theorem\cite{Mermin1966}.
To this end, diagrammatic extensions of DMFT\cite{Rohringer2018a} have been developed where a Feynman diagrammatic expansion around a correlated DMFT starting point is performed by constructing Feynman diagrams from the local vertex and nonlocal Green's functions of DMFT.
In the last decade this idea has been used in various ways which led to the development of the dynamical vertex approximation (D$\Gamma$A)\cite{toschi2007a} in its parquet\cite{Valli2015,Kauch2022} and ladder\cite{Katanin2009,Held2008,DelRe2021a} implementation, the dual fermion\cite{Rubtsov2008,Rubtsov2009,Krien2020a,Hirschmeier2015} and dual boson\cite{Rubtsov2012} methods, the one-particle irreducible approach\cite{Rohringer2013} as well as the TRILEX\cite{Ayral2015,Ayral2016a} and QUADRILEX\cite{Ayral2016} methods.
Within the ladder D$\Gamma$A nonlocal correlations sizably reduce the DMFT transition temperature to an antiferromagneticlly ordered state in the three-dimensional bipartite Hubbard model\cite{Rohringer2011,Rohringer2016} and lead to a suppression of the magnetically ordered state at any finite temperature on the two-dimensional square lattice in accordance with the Mermin-Wagner theorem\cite{Schafer2015}.

In this paper, we extend these ladder D$\Gamma$A studies of the three-dimensional Hubbard model on a simple cubic lattice to various three-, four- and five-dimensional lattices to gain insights in the specific role played by lattice coordination and dimension on the magnetic transition.
More specifically, we consider the body- and face-centered-cubic lattices in three dimensions as well as the simple (hyper)cubic lattices in three, four (tesseract), and five dimensions and compare their  magnetic DMFT phase diagram for selected fillings to corresponding ladder D$\Gamma$A results.
%\sout{This allows us to quantify the impact of nonlocal correlations and, in particular, its dependence on the lattice geometry.}
We find that, in general, the difference between DMFT and ladder D$\Gamma$A decreases with increasing coordination number and/or dimension which is indeed the expected behavior, as DMFT is known to perform better in higher dimensions (or, more generally, for systems with higher coordination number) and eventually becomes exact in the limit of infinite coordination number\cite{Muller-Hartmann1989}.
%\gr{I would replace Ale's suggestion for the next sentence by the following: 
Moreover, in four and five dimensions we simultaneously obtain, as expected, both a sizable reduction of $T_N$ in lD$\Gamma$A with respect to DMFT as well as critical exponents compatible with the mean-field (MF) universality class (except for logarithmic corrections in four dimensions) which supports the overall reliability of the approach to study magnetic transitions of correlated electrons beyond MF/DMFT.
%\at{Further, we will also show how, when applying the  D$\Gamma$A in four dimensions, one obtains simultaneously, as it should,  both a sizable reduction w.r.t~DMFT of the ordering temperature of the magnetic transitions, as well as critical exponents compatible with the MF-universality class, albeit logarithmic corrections, which supports the overall reliability of the approach to study magnetic transitions of correlated electrons beyond MF/DMFT.}

Apart from describing these general trends, we provide accurate values for the transition temperatures in the strong coupling Heisenberg regime, and compare them with the corresponding DMFT values. This can serve as an insightful guidance for a quick estimate (or prediction) of experimentally observed critical temperatures for different material geometries, based on purely DMFT results (e.g., in the situations where realistic calculations beyond DMFT are not feasible), as well as a reliable benchmark for future  studies of the magnetically ordered phases of strongly correlated electrons.

Finally, we quantify the importance of nonlocal correlations by calculating the magnitude of the violation of specific sum rules within DMFT which originate from its inability to capture spatial fluctuations with the necessary accuracy. 

The paper is organized as follows: In Sec.~\ref{sec:model} we introduce the Hubbard model on the different lattices mentioned above and give a short overview about the ladder D$\Gamma$A approximation. In Sec.~\ref{sec:results}, we present our results for the spin susceptibilities of both DMFT and ladder D$\Gamma$A, the critical exponent as well as the magnetic phase diagram for DMFT and D$\Gamma$A. We also discuss the phase transition at weak coupling and analyze how the violation of a specific sum rule within DMFT depends on the geometric structure of the lattice. Finally, in Sec.~\ref{sec:conclusions} we summarize our results and discuss future research perspectives. In appendix~\ref{sec:rpa} we present random phase approximation (RPA) results for the fcc lattice.

%The effects of correlation in correlated electron systems are strongly influenced by the geometry of the underlying lattice\cite{Rohringer2011,Schafer2015}. 

\section{Model and Method}
\label{sec:model}

\begin{figure}
    \includegraphics[width = \textwidth]{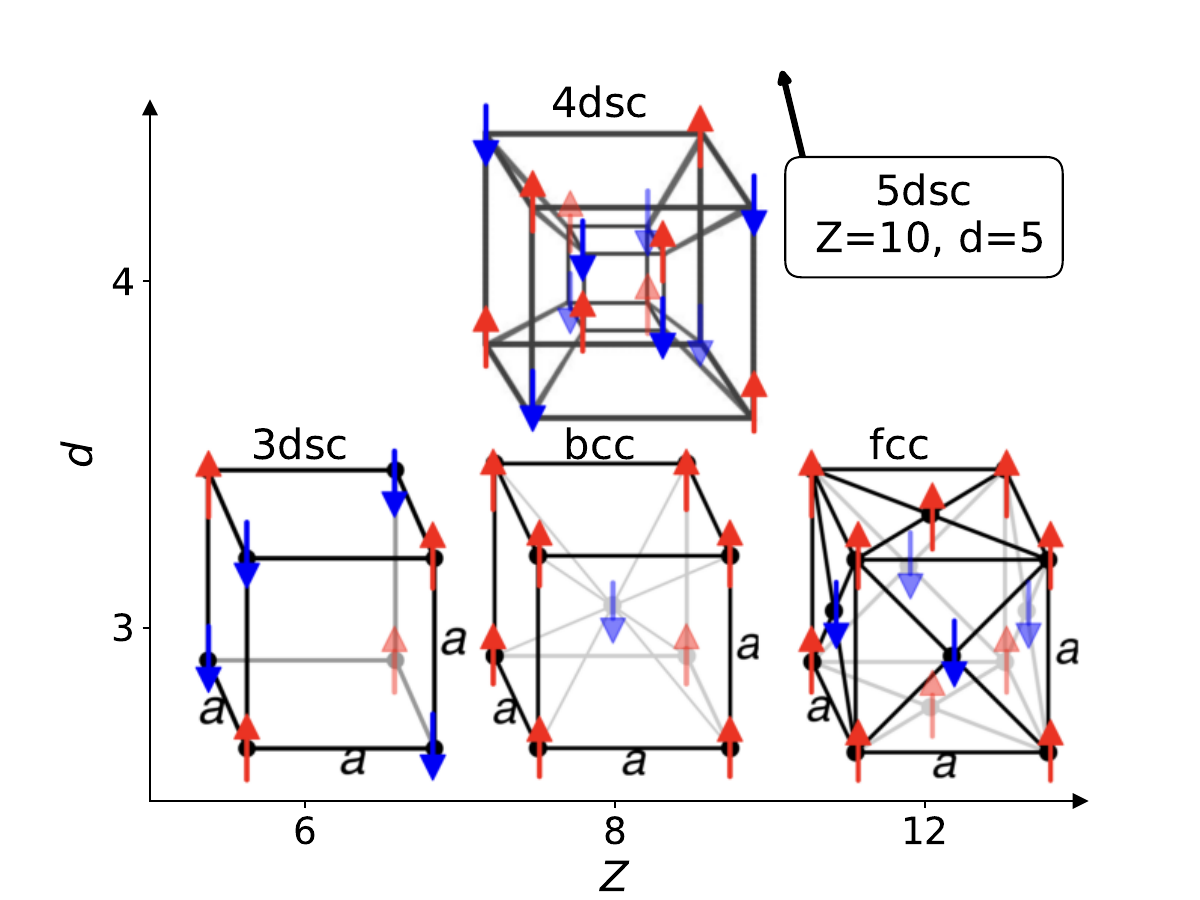}
    \caption{Three dimensional simple cubic (3d-sc), body-centered-cubic (bcc), face-centered-cubic (fcc) and four dimensional simple cubic (4d-sc, tesseract) lattices considered in this study ordered by increasing coordination number $Z$ (from left to right) and increasing spatial dimension $d$ (from bottom to top). Since projections of the five dimensional cubic lattice are barely recognizable on paper, it is only signified by its coordination number and dimension. Arrows display the spin arrangement for the predominant magnetic order.}
    \label{fig:lattice_overview}
\end{figure}

For this study we consider the single-band Hubbard model
\begin{equation}
H = -t\sum_{\sigma \langle ij\rangle} c_{i\sigma}^\dagger c_{j\sigma} + U \sum_i n_{i\uparrow}n_{i\downarrow}
\label{eq:hamiltonian}
\end{equation}
where $c^{(\dagger)}_{i\sigma}$ annihilates (creates) an electron with spin $\sigma$ at lattice site $\mathbf{R}_i$, $n_{i\sigma}\!=\!c^{\dagger}_{i\sigma}c_{i\sigma}$ is the particle number, $t$ denotes the hopping amplitude between nearest neighbors and $U$ is the on-site Hubbard interaction.
We will study this model on the following three-, four-, and five-dimensional lattices which are shown in Fig.~\ref{fig:lattice_overview} with increasing coordination number from left to right and increasing dimension from bottom to top: the simple-cubic lattice in three dimensions (lower row, left), the body-centered-cubic lattice (lower row, middle), the face-centered-cubic lattice (lower row, right) and the simple hypercubic lattice (upper row, middle) in four dimensions (tesseract), while the five dimensional hypercubic lattice is only indicated by its coordination number.
%\gr{
Let us note that for the representation of the $4d$ hypercubic lattice we have selected a projection to three dimensions where the coordination number of the lattice can be seen in the most transparent way\cite{Coxeter1973,Conway2008}.
%}
We choose the hopping parameter $t$ to fix the second moment of the noninteracting density of states to $0.5$ which allows for a direct comparison between the different lattices in DMFT\cite{Bulla2000}. 
This implies that $t=1/\left(2\sqrt{Z}\right)$ where $Z$ is the coordination number of the lattice (i.e., $Z_\text{bcc} \! = \! Z_\text{4d-sc} \! = \! 8$, $Z_\text{5d-sc} \! = \! 10$, $Z_\text{fcc} \! = \! 12$ to be compared with $Z_\text{3d-sc} =6$). 
The tight-binding dispersion relations for each lattice are given by:
\begin{subequations}
\label{equ:dispersionrelations}
\begin{align}
\epsilon_{\mathrm{3d-sc}} =& -2t\left(\cos ak_x +\cos ak_y +\cos ak_z \right)  \label{eq:3d_disp}\\
\epsilon_{\mathrm{bcc}} =& -8t\cos \frac{ak_x}{2} \cos \frac{ak_y}{2} \cos \frac{ak_z}{2}  \\
\epsilon_{\mathrm{fcc}} =& -4t\left(\cos\frac{ak_x}{2}\cos\frac{ak_y}{2} +\cos\frac{ak_x}{2}\cos\frac{ak_z}{2}\right. \label{eq:bcc_disp}\\
&\left.+\cos\frac{ak_y}{2}\cos\frac{ak_z}{2}\right) \nonumber\label{eq:fcc_disp} \\
\epsilon_{\mathrm{4d-sc}} =& -2t\left(\cos ak_1 +\cos ak_2  \right.\\
&\left.+\cos ak_3 +\cos ak_4 \right)\\
\label{eq:4d_disp}
\epsilon_{\mathrm{5d-sc}} =& -2t\left(\cos ak_1 +\cos ak_2  \right.\\
&\left.+\cos ak_3 +\cos ak_4 + \cos ak_5 \right)\nonumber,
\label{eq:5d_disp}
\end{align}
\end{subequations}
where we set $a\!=\!1$.
In the following, we denote fermionic Matsubara frequencies with $\nu_n = (2n+1)\frac{\pi}{\beta}$ and bosonic Matsubara frequencies with  $\omega_n = 2n\frac{\pi}{\beta}$, where $\beta = \frac{1}{T}$ is the inverse temperature and $n\in\mathds{Z}$. 
Finally we indicate Matsubara sums as well as integrals over the Brillouin zone as $\sum_\nu := \frac{1}{\beta} \sum_{\nu}$ and $\sum_k := \frac{1}{V_{\mathrm{BZ}}}\int_{V_{\mathrm{BZ}}}\mathrm{d}\mathbf{k}$, respectively.\\
%This study makes use of DMFT and its diagramatic extension D$\mathrm{\Gamma}$A.

\subsection{Ladder D\texorpdfstring{$\mathbf{\Gamma}$}{G}A}
\label{sec:DMFTDGA}

Since the Hubbard model in Eq.~\eqref{eq:hamiltonian} is not exactly solvable for the three-, four-, and five-dimensional lattices considered here, we have to resort to approximate approaches.
In a first step, we will exploit DMFT to calculate one- and two-particle correlation functions, in order to capture all purely local correlations of the system.
Within this approach, the single-particle Green's function $G^\nu_\mathbf{k}$ is obtained as

\begin{equation}
\label{equ:DefDMFTGF}
G_\mathbf{k}^\nu=\frac{1}{i\nu+\mu-\epsilon_\mathbf{k}-\Sigma_\text{imp}^\nu}
\end{equation}
where $\Sigma_\text{imp}^\nu$ 
%and $G_\text{imp}^\nu$ are 
is the local single particle self-energy 
%and local Green's function 
of an auxiliary single-impurity Anderson model (SIAM) and $\mu$ denotes the chemical potential which fixes the electron density of the system.
For the bipartite 3d-sc, bcc, 4d-sc, and 5d-sc lattices, we consider $\mu\!=\!\frac{U}{2}$ which enforces half filling $n\!=\!1$, while for the strongly frustrated fcc lattice $\mu$ has to be determined to obtain the required filling.
The SIAM from which the local self-energy is obtained has to fulfill the following self-consistency condition
\begin{equation}
\label{equ:selfconst}
\sum_\mathbf{k}\underset{G_\mathbf{k}^\nu}{\underbrace{\frac{1}{i\nu+\mu-\epsilon_\mathbf{k}-\Sigma_\text{imp}^\nu}}}\overset{!}{=}\underset{G_\text{imp}^\nu}{\underbrace{\frac{1}{i\nu+\mu-\Delta_\text{imp}^\nu-\Sigma_\text{imp}^\nu}}},
\end{equation}
where $\Delta_\text{imp}^\nu$ is the hybridaztion function which defines the SIAM and $G_\text{imp}^\nu=[i\nu+\mu-\Delta_\text{imp}^\nu-\Sigma_\text{imp}^\nu]^{-1}$ represent the local impurity Green's function.

The DMFT spin (magnetic) susceptibility $\chi_{m,\mathbf{q}}^\omega$ is obtained by summing the generalized susceptibility over the fermionic Matsubara frequencies
\begin{equation}
\label{equ:physical_susceptibility}
\chi_{m,\mathbf{q}}^{\omega} = \sum_{\nu\nu'} \chi_{m,\mathbf{q}}^{\nu\nu'\omega}
\end{equation}
where $\chi_{m,\bq}^{\nu\nu'\omega}$ can be calculated from the DMFT Green's function $G_\mathbf{k}^\nu$ in Eq.~\eqref{equ:DefDMFTGF} and the local irreducible spin vertex $\Gamma_m^{\nu\nu'\omega}$  via the Bethe-Salpeter (BS) equation
\begin{equation}
\label{equ:generalizedsusc}
\chi_{m,\mathbf{q}}^{\nu\nu'\omega} = \chi_{0,\mathbf{q}}^{\nu\nu'\omega} - 
\sum_{\nu_1\nu_2} \chi_{0,\bq}^{\nu\nu_1\omega} \Gamma_{m}^{\nu_1\nu_2\omega}\chi_{m,\bq}^{\nu_1\nu'\omega},
\end{equation}
where $\chi_{0,\mathbf{q}}^{\nu\nu'\omega}=-\beta\sum_\mathbf{k}G_\mathbf{k}^\nu G_\mathbf{k+q}^{\nu+\omega}\delta_{\nu\nu'}$ is the so-called bubble contribution to the susceptibility which does not include vertex corrections.

As a mean-field theory with respect to spatial degrees of freedom DMFT typically overestimates ordered states which is reflected in an overestimation of the corresponding spin susceptibility $\chi_{m,\mathbf{q}}^\omega$.
This problem can be mitigated by taking into account the contribution of nonlocal correlations within the D$\Gamma$A. 
Within this approach the mutual renormalization between nonlocal fluctuations in the charge, spin and particle-particle scattering channels are taken into account by means of the parquet equations~\cite{Yang2009,Tam2013,Valli2015,Li2016,Li2017,Eckhardt2020,Krien2020,Astretsov2020}, which construct all one- and two-particle correlation functions and susceptibilities from a single input, i.e., the fully irreducible two-particle vertex $\Lambda_{\text{imp},\sigma\sigma'}^{\nu\nu'\omega}$ of the auxiliary SIAM. 
Note that the frequency dependence of this vertex function directly encodes the highly nonperturbative\cite{Pelz2023} physics of electronic localization \cite{Gunnarsson2017,Reitner2020, Chalupa2021,Adler2024} associated with the Mott transition as well as strong quasiparticle renormalization effects.
While this procedure would certainly provide the most accurate results, it is limited to rather high temperatures above the transition to magnetically ordered phases due to its considerable numerical complexity and, hence, not suitable for the present study.
We, therefore, stick to the ladder approximation of D$\Gamma$A (lD$\Gamma$A) where $\chi_{m,\mathbf{q}}^\omega$ is evaluated via Eqs.~\eqref{equ:physical_susceptibility} and \eqref{equ:generalizedsusc} which requires only a solution of a single BS equation in one channel.
On the down side, this procedure leads to a violation of an exact sum rule (which  would be instead automatically fulfilled within the parquet technique) given by
\begin{equation}
    \frac{1}{2}\sum_{\omega\mathbf{q}}\left(\chi_{d,\mathbf{q}}^\omega + \chi_{m,\mathbf{q}}^\omega\right) \overset{!}{=} \frac{n}{2}(1-\frac{n}{2})
    \label{eq:pauli_principle}
\end{equation}
where $\chi_{d,\mathbf{q}}^\omega$ is the charge susceptibility which can be obtained in a completely analogous way as the corresponding spin susceptibility in Eqs.~\eqref{equ:physical_susceptibility} and \eqref{equ:generalizedsusc}.
To restore this sum rule within our ladder approach we apply an effective correction to the DMFT spin susceptibility via a constant parameter $\lambda$ which transforms the DMFT to the lD$\Gamma$A spin susceptibility
\begin{equation}
    \chi_{m,\mathbf{q}}^{\lambda,\omega}= \frac{1}{\left(\chi_{m,\mathbf{q}}^\omega\right)^{-1} + \lambda}.
    \label{eq:moriya}
\end{equation}
This idea has been first suggested by Moriya in his work on itinerant magnetism\cite{Moriya1985a} and later exploited in the ladder D$\Gamma$A\cite{Katanin2009}. 
The value of $\lambda$ is determined to satisfy the sum rule in Eq.~\eqref{eq:pauli_principle}.
This $\lambda$ correction has a clear physical interpretation: Close to a phase transition the static ($\omega\!=\!0$) spin susceptibility takes an Ornstein-Zernike\cite{Ornstein1914} form

\begin{equation}
\label{equ:OrnsteinZernike}
\chi_{m,\mathbf{q}}^{\omega=0}=\frac{A}{(\mathbf{q}-\mathbf{q}_N)^2+\xi^{-2}},
\end{equation}
where $\mathbf{q}_N$ is the ordering vector (equivalent to the nesting vector in bipartite lattices), $\xi$ is the correlation length and $A$ is a (temperature dependent) parameter. 
Within this picture, the $\lambda$ correction corresponds to a simple renormalization of the correlation length, which is overestimated within DMFT and reduced by the effective inclusion of nonlocal correlations and mutual renormalization via the $\lambda$ parameter.

\section{Numerical Results}
\label{sec:results}

\begin{figure}
    \includegraphics[width = \textwidth]{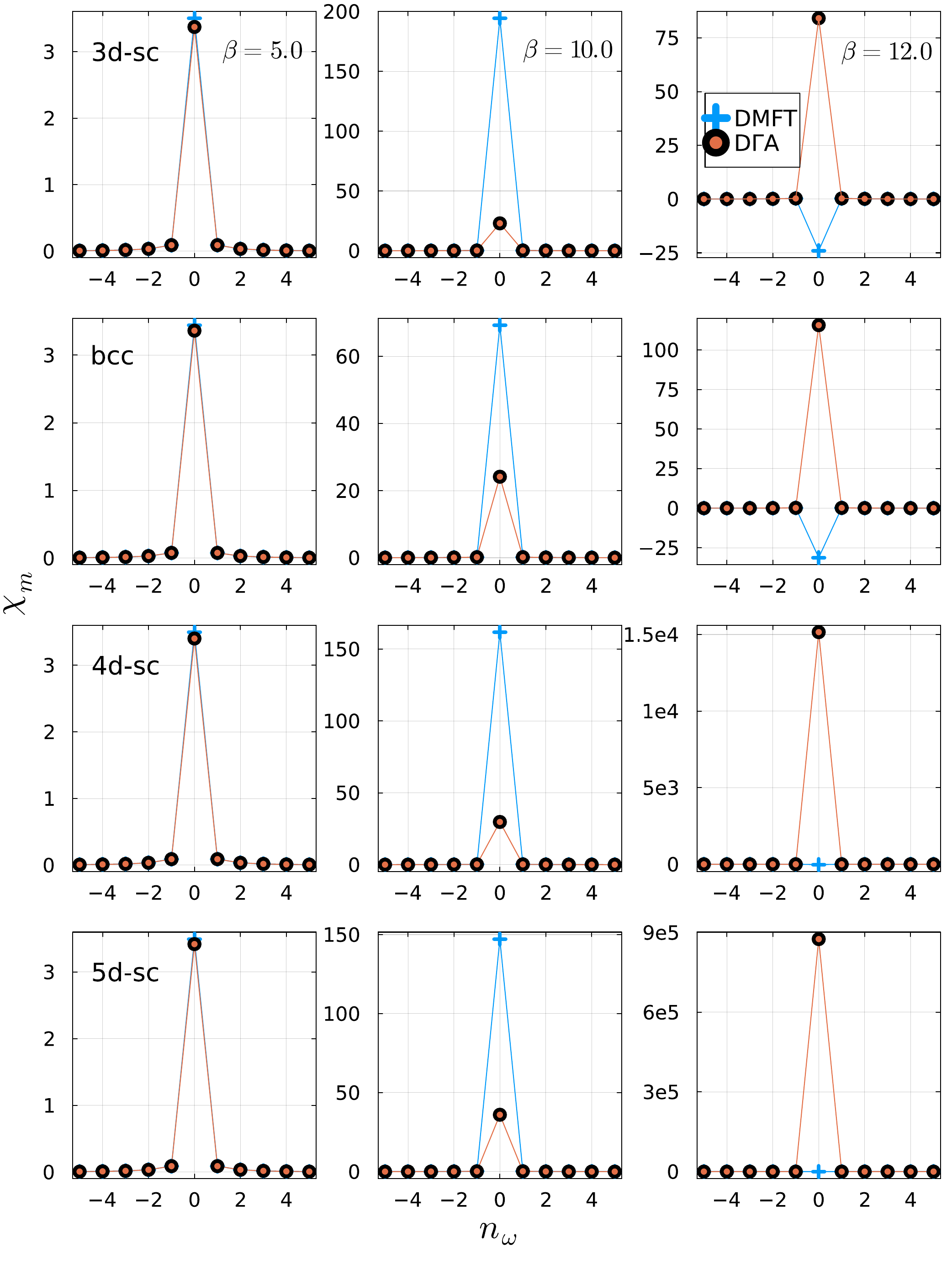}
    \caption{Spin susceptibilities $\chi_{m,\mathbf{q}=\mathbf{q}_N}^\omega$ at $U=2.0$ and $n=1.0$ for the three temperatures $\beta = 5.0$ (left panels), $\beta=10.0$ (middle panels) and $\beta=12.0$ (right panels) for the 3d-sc (first row), bcc (second row), 4d-sc (third row) and 5d-sc (fourth row) lattices, respectively. 
    Data are shown for DMFT (blue crosses) and lD$\Gamma$A (orange circles). The nesting vectors (which correspond to antiferromagnetic ordering vectors for bipartite lattices) are $\mathbf{q}_N=(\pi,\pi,\pi)$ (3d-sc), $\mathbf{q}_N=(2\pi,2\pi,2\pi)$ (bcc), $\mathbf{q}_N=(\pi,\pi,\pi,\pi)$ (4d-sc) and $\mathbf{q}_N=(\pi,\pi,\pi,\pi,\pi)$ (5d-sc).
    % Each row corresponds to one lattice type with decreasing temperatures (corresponding to increasing $\beta$) from left to right. 
    %\gr{Maybe you can put the $\beta$ value into the respective upper panels. Moreover, the symbols really have to be larger otherwise they are difficult to differentiate.}
    %To keep compatibility each plot has been normalized to its corresponding $\chi^{\omega = 0}_{m,\mathrm{D\Gamma A}}$ (abbreviated as $\chi_m^{\omega_0}$ for readabilities sake).
    }
    \label{fig:bipartite_susc}
\end{figure}

\begin{figure*}[t!]
    \includegraphics[width = 0.24\textwidth]{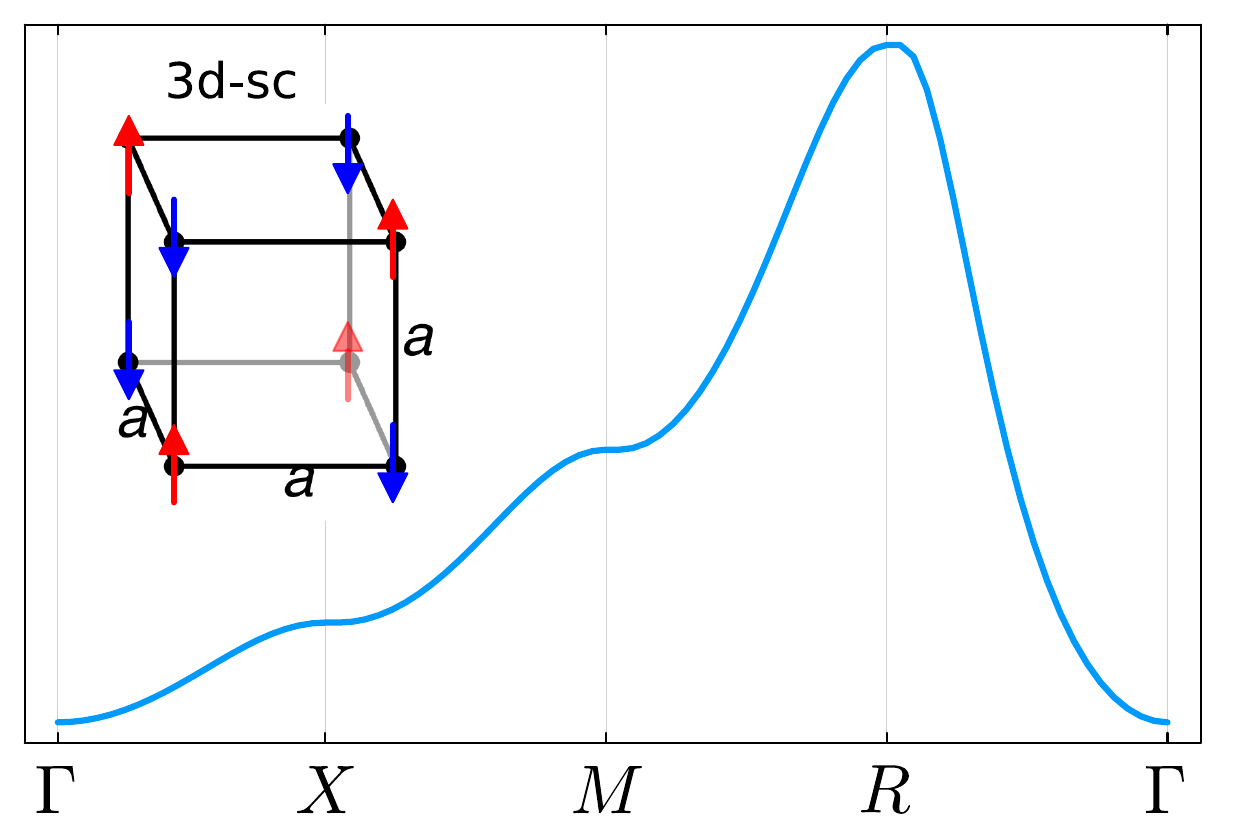}
    \includegraphics[width = 0.24\textwidth]{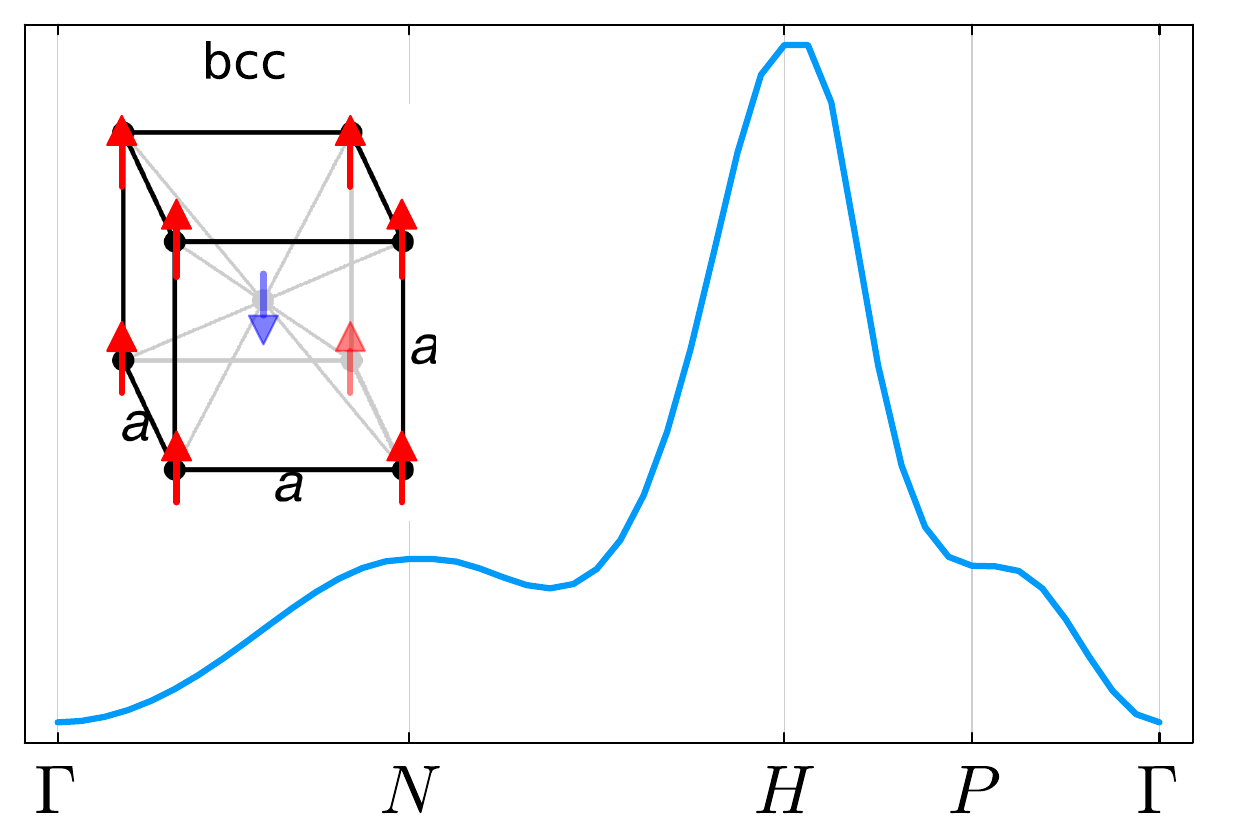}
    \includegraphics[width = 0.24\textwidth]{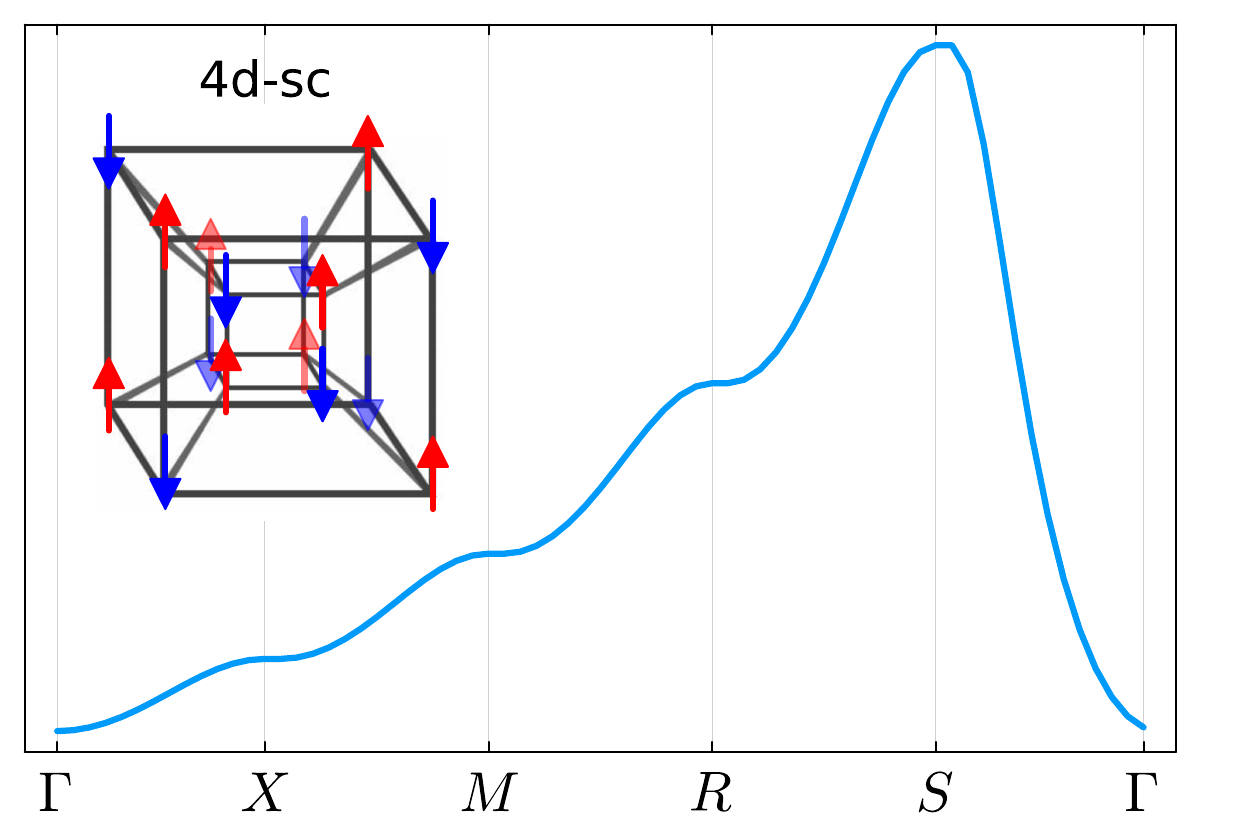}
    \includegraphics[width = 0.24\textwidth]{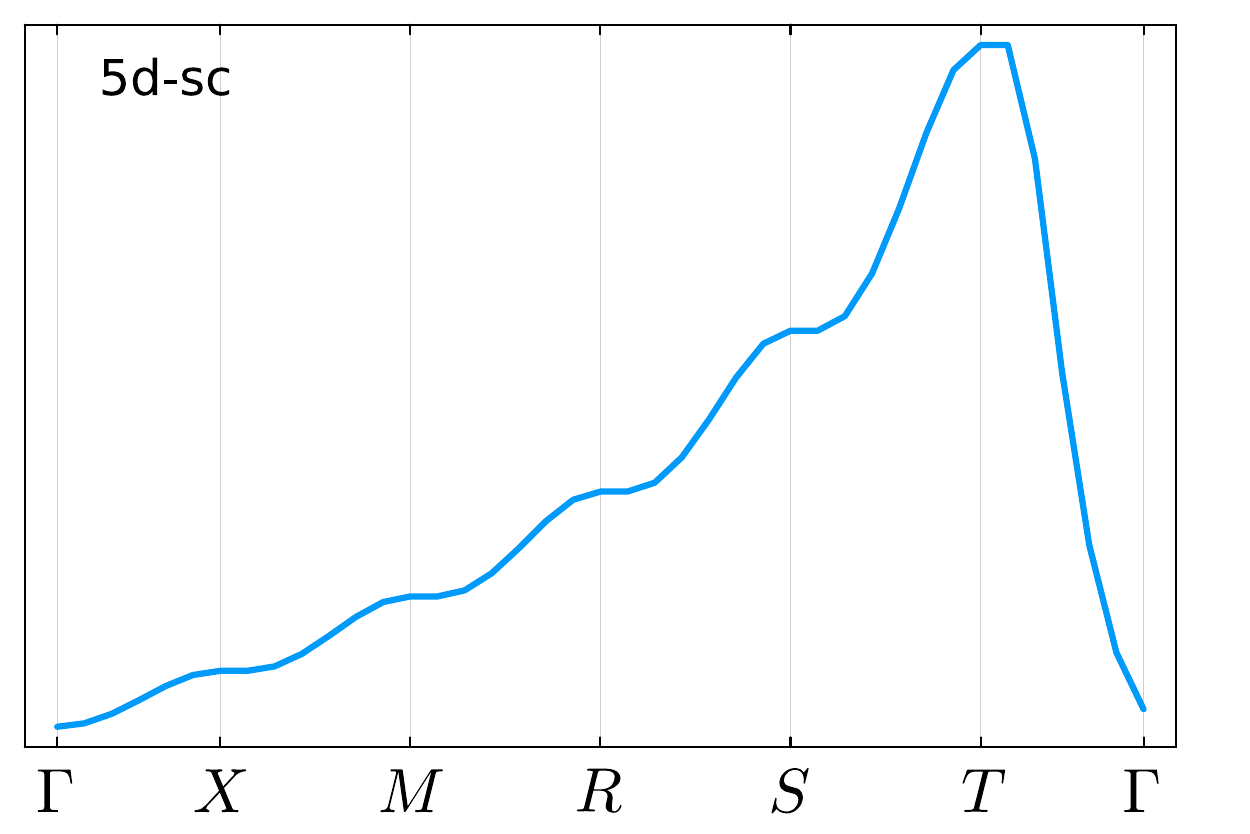}
    \caption{%\gs{For me the spins in the sketches of the various lattices are too small. Also, the 4d-lattice has a different orientation w.r.t. other three....not so nice. Can this be changed? I would also remove the ytics (gnuplot-language) in these plots.} 
    Momentum dependence  of the static DMFT susceptibilities $\chi_{m,\mathbf{q}}^{\omega=0}$ along a high-symmetry path for the 3d-sc (first panel), the bcc (second panel), the 4d-sc (third panel) and the 5d-sc (fourth panel) lattice, respectively, at $U=2.0$, $n=1.0$ and $\beta = 5.0$.  Since there is no established convention for labeling high symmetry points in the 4d-sc and 5-sc lattice, we choose a notation similar to the 3d-sc lattice: $\Gamma = (0,0,0,0)$, $X = (0,0,0,\pi)$, $M = (0,0,\pi,\pi)$, $R = (0,\pi,\pi,\pi)$, $S = (\pi,\pi,\pi,\pi)$ for the 4d-sc lattice and $\Gamma = (0,0,0,0,0)$, $X = (0,0,0,0,\pi)$, $M = (0,0,0,\pi,\pi)$, $R = (0,0,\pi,\pi,\pi)$, $S = (0,\pi,\pi,\pi,\pi)$ and $T = (\pi,\pi,\pi,\pi,\pi)$ for the 5d-sc lattice. Insets: geometrical structure of the lattices replotted from Fig.~\ref{fig:lattice_overview} where the arrows represent the predominant spin order.}
    %Each lattice structure shows its anti-ferromagnetic state, represented by red up arrows and blue down arrows. 
    %In the fcc lattice, the state presented here is given by the ordering vector $\mathbf{q} = (0,0,2\pi)$.% \gr{You could at the label indicating the corresponding lattic type in the figures.}}
    %}
    \label{fig:qpaths}
\end{figure*}

In the following, we discuss our numerical results for the five lattices under consideration.
The 3d-sc, bcc, 4d-sc and 5d-sc lattices are bipartite, which implies their perfect particle-hole symmetry at half filling ($n\!=\!1$ particle per lattice site.)
In this situation, the Fermi surface is independent of the interaction strength $U$ as well as of the temperature $T$, and features perfect nesting, i.e., for each point $\mathbf{k}$ on the Fermi surface $\mathbf{k}+\mathbf{q}_N$ is also located on the Fermi surface where $\mathbf{q}_N$ is the so-called nesting vector.
The latter property entails a strong tendency to magnetic order at wave vector $\mathbf{q}_N$ indicated by a large (static) spin-susceptibilities $\chi_{m,\mathbf{q}_N}^{\omega=0}$.
As we are particularly interested in the corresponding large spin fluctuations above the transition temperature to the ordered state as a function of the interaction strength $U$ we will perform our analysis exactly at half filling.

The fcc lattice, on the other hand, is strongly geometrically frustrated which leads to a substantial suppression of magnetic order.
%\gr{
This has been indeed observed in real materials with a fcc lattice structure such as Ba$_2$LnSbO$_6$ and Sr$_2$LnSbO$_6$ double perovskites (with Ln=Dy, Ho, Gd) where no magnetic order is found down to $T=2$ K~\cite{Karunadasa2003}, in the molecular antiferromagnet fcc-Cs$_3$C$_{60}$ where $T_N=2.2$ K~\cite{Kasahara2014}, and in Ba$_2$YbTaO$_6$ where no magnetic long-range order is found down to $T=100$ mK~\cite{Sahu2025}.
Let us, however, note that a next-nearest neighbor hopping in the fcc lattice, which favors a type-II antiferromagnetic order where ferromagnetic layers are antiferromagnetically stacked along the [111] direction and spins at neighboring corners of the cube are pointing in opposite direction, can again enhance magnetic order as it is for instance the case in NiO\cite{Chatterji2009}.
On the theoretical side, the suppression of antiferromagnetic order is associated with the absence of  particle-hole symmetry where even at half filling no nesting vector exists which would enhance spin fluctuations.
%}
In fact, at exactly $n\!=\!1$ RPA and slave boson methods\cite{Igoshev2015} predict that the system is at the verge of a number of competing magnetic phases (including also regions of phase separation) at $T\!=\!0$ which makes our numerical analysis challenging also at finite temperatures (see the RPA analysis in appendix~\ref{sec:rpa}).
Hence, for studying the magnetic transition of the fcc lattice, we have selected $n\!=\!0.75$ around which a stable ordered ground sate with the ordering vector $\mathbf{q}_N\!=\!(0,0,2\pi)$ is observed in RPA (see appendix~\ref{sec:rpa}).
However, since the strong geometric frustration makes the fcc lattice rather different from the four others, we will discuss the latter together and address the fcc lattice in a separate paragraph for each observable/correlation function which we present in the following.

\subsection{Frequency and momentum dependence of spin susceptibilities}% and Correlation Lengths}
\label{sec:susclength}

In this section, we analyze the frequency and momentum dependence of the spin susceptibilities for different temperatures and interaction strengths for both the bipartite and the fcc lattices.

{\sl Bipartite lattices:} In Fig.~\ref{fig:bipartite_susc}, the spin susceptibilities $\chi_{m,\mathbf{q}=\mathbf{q}_N}^\omega$ at the nesting vector $\mathbf{q}=\mathbf{q}_N$ (see caption of Fig.~\ref{fig:bipartite_susc} are compared between DMFT (blue crosses) and lD$\Gamma$A (orange circles) as a function of the bosonic Matsubara frequency $\omega$ for the 3d-sc, bcc, 4d-sc and 5d-sc lattices in the first, second, third and fourth row, respectively, at $\beta=5.0$ (left panels), $\beta=10.0$ (middle panels) and $\beta=12.0$ (right panels) for $U\!=\!2.0$.
Let us note, that at the nesting (ordering) vector $\mathbf{q}\!=\!\mathbf{q}_N$ the largest difference between DMFT and lD$\Gamma$A is expected.
%The corresponding D$\Gamma$A results are indicated by  a orange line. 
% Note that the temperatures for which we present our data slightly differ for the three different lattices as the correspond to the same distances to the transition temperature $T_N$ which is different for the three lattices.
The spin susceptibilities are always strongly peaked at $\omega\!=\!0$ as classical spin fluctuations are typically predominant.
At the largest temperature $\beta=5.0$ (left panels in Fig.~\ref{fig:bipartite_susc}), our D$\Gamma$A results almost coincide with the DMFT data, which is consistent with the fact that DMFT becomes the exact solution of the Hubbard model also in the limit of infinite temperatures (where the system becomes effectively noninteracting as the energy scale of the interaction becomes negligible compared to the temperature). %\at{[AT: haben wir eine Referenz für diese Aussage?]}.
%\gr{Ich habe keine gefunden, aber kann es anders sein? Ween $T>>>U$ dann haben wir ja wieder ein nichtwechselwirkendes System?}
Upon decreasing the temperature and approaching the DMFT phase transition to the magnetically ordered state, we observe that both DMFT and D$\Gamma$A spin susceptibilities are enhanced, whereas the DMFT one increases much faster (middle panels).
This difference between DMFT and D$\Gamma$A data is a consequence of the renormalization of $\chi_{m,\mathbf{q}}^\omega$ within the lD$\Gamma$A to fulfill the sum rule in Eq.~\eqref{eq:pauli_principle}.
At the lowest temperatures (right panels) DMFT already predicts an ordered phase indicated by the negative spin susceptibility at $\omega\!=\!0$ which is turned to a large but positive value via the $\lambda$ correction within D$\Gamma$A for which, instead, we are still in the paramagnetic regime.
Note that the absolute values of $\chi_{m,\mathbf{q}=\mathbf{q}_N}^\omega$ are quite different between the four lattice types, due to the difference in their respective DMFT and D$\Gamma$A transition temperatures.

In Fig.~\ref{fig:qpaths} we depict the static spin susceptibility $\chi_{m,\mathbf{q}}^{\omega=0}$ as function of the momentum $\mathbf{q}$ along a high-symmetry path in the Brillouin zone for $U\!=\!2.0$ for the 3d-sc, bcc, 4d-sc and 5d-sc lattices, respectively. 
We present only DMFT data here as the functional form of $\chi_{m,\mathbf{q}}^{\omega=0}$ cannot be modified by the momentum independent $\lambda$ correction of lD$\Gamma$A and the overall reduction in size due to this correction has already been discussed for the local spin susceptibilities in Fig.~\ref{fig:bipartite_susc}.
Even for the rather high temperature $\beta\!=\!5.0$, we observe a clear peak of these correlation functions at the respective ordering (nesting) vector $\mathbf{q}_N$ for all four bipartite lattices.
This demonstrates that spin fluctuations are strongly pronounced already quite far away from the low-temperature phase transition.

\begin{figure}[t!]
    \includegraphics[width = \textwidth]{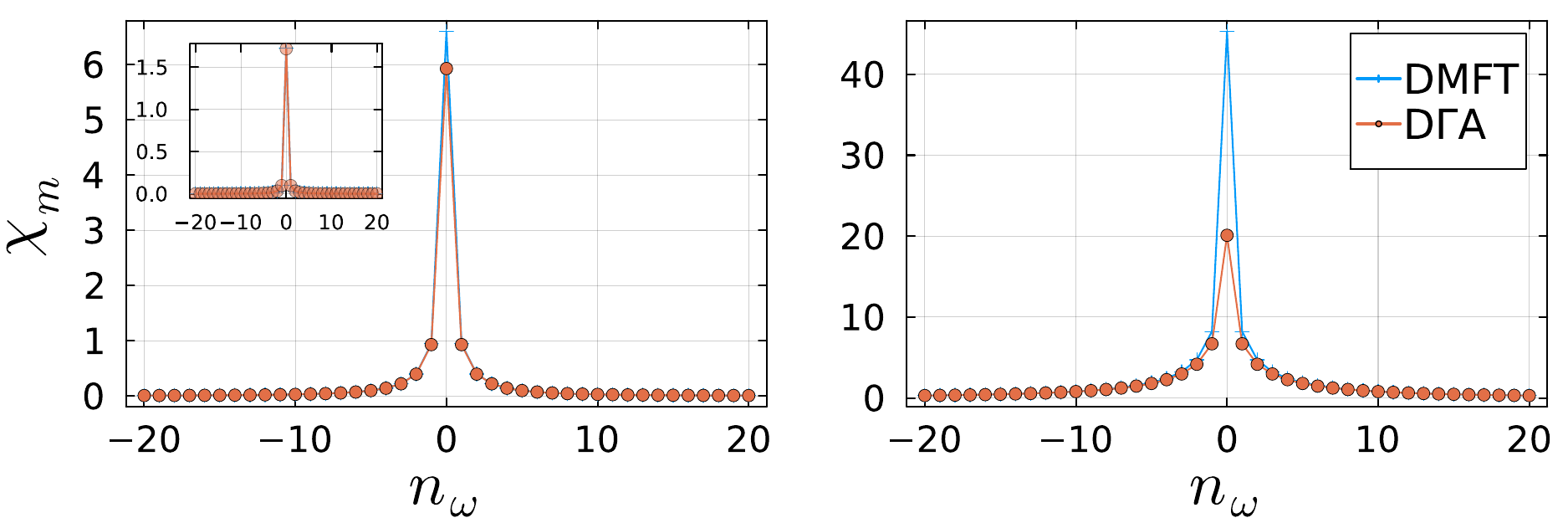}
    \caption{Spin susceptibility $\chi_{m,\mathbf{q}=\mathbf{q}_N}^\omega$ for the fcc lattice at $U=3.0$ and $n=0.75$ for $\beta=5.0$ (inset of left panel), $\beta = 20.0$ (left panel), and $\beta = 150.0$ (right panel). The ordering vector $\mathbf{q}_N=(0,0,2\pi)$ corresponds to the layered antiferromagnetic order depicted in Fig.~\ref{fig:fcc-susc-q}.} 
    %at the ordering vector $\mathbf{q}_N = (0 ,0,2\pi)$.}
    %$\beta = 20.0$ ($\lambda = 0.01714752$), $\beta = 150.0$ ($\lambda=0.02766269$)
    \label{fig:fcc-susc}
\end{figure}

\begin{figure}[t!]
    \includegraphics[width = \textwidth]{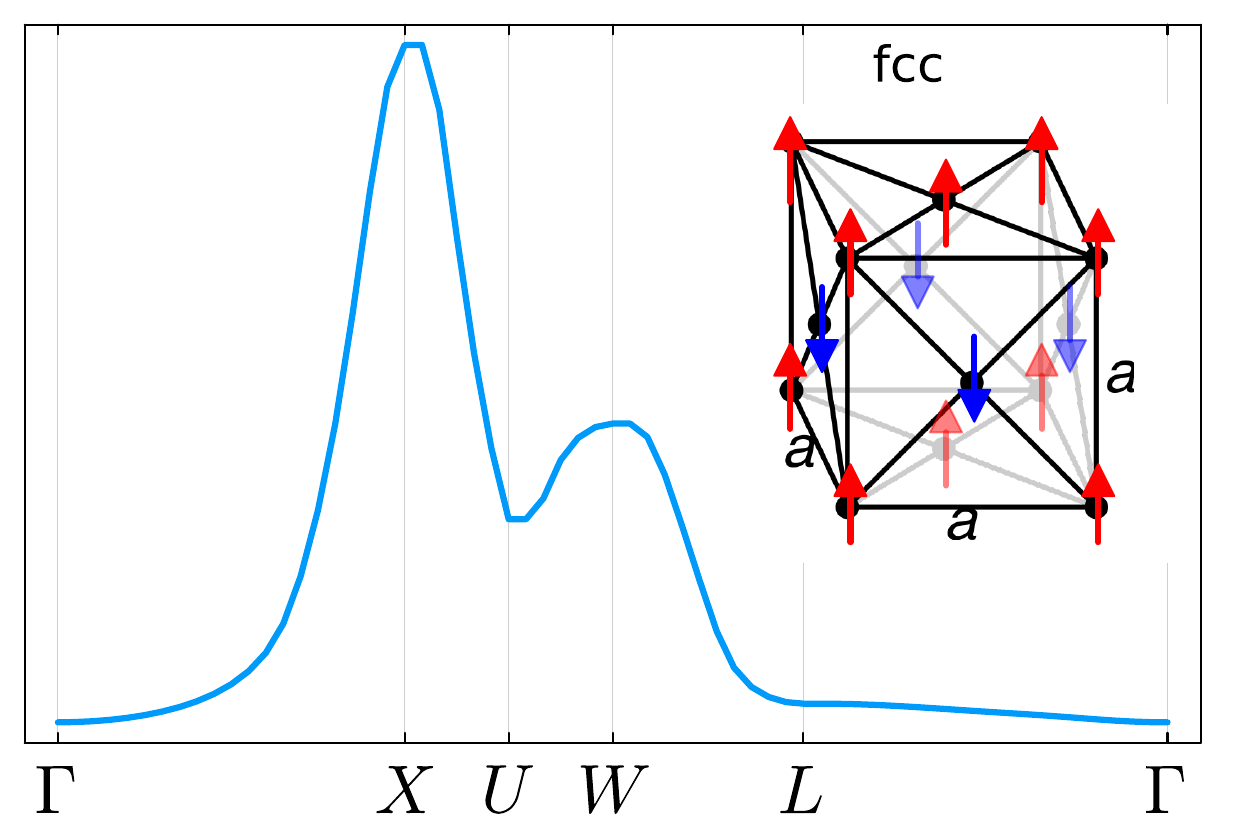}
    \caption{ Momentum dependence  of the static DMFT susceptibilities $\chi_{m,\mathbf{q}}^{\omega=0}$ along a high-symmetry path for the fcc lattice at $U=3.0$, $n=0.75$ and $\beta = 80.0$. The spin-ordered state represented by the arrows corresponds to the ordering vector $\mathbf{q} = (0,0,2\pi)$.}
    %$\beta = 20.0$ ($\lambda = 0.01714752$), $\beta = 150.0$ ($\lambda=0.02766269$)
    \label{fig:fcc-susc-q}
\end{figure}

\begin{figure*}[t!]
    \includegraphics[width = 0.48\textwidth]{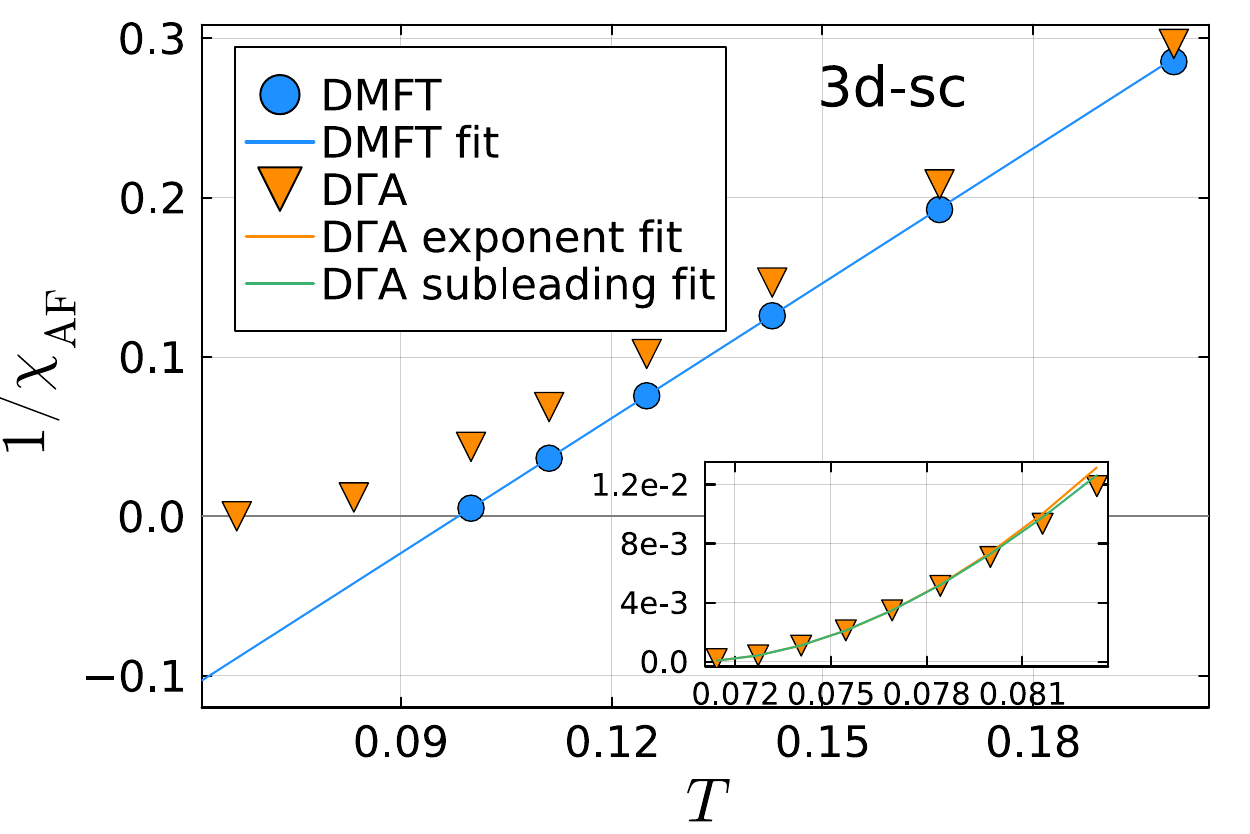}
    \includegraphics[width = 0.48\textwidth]{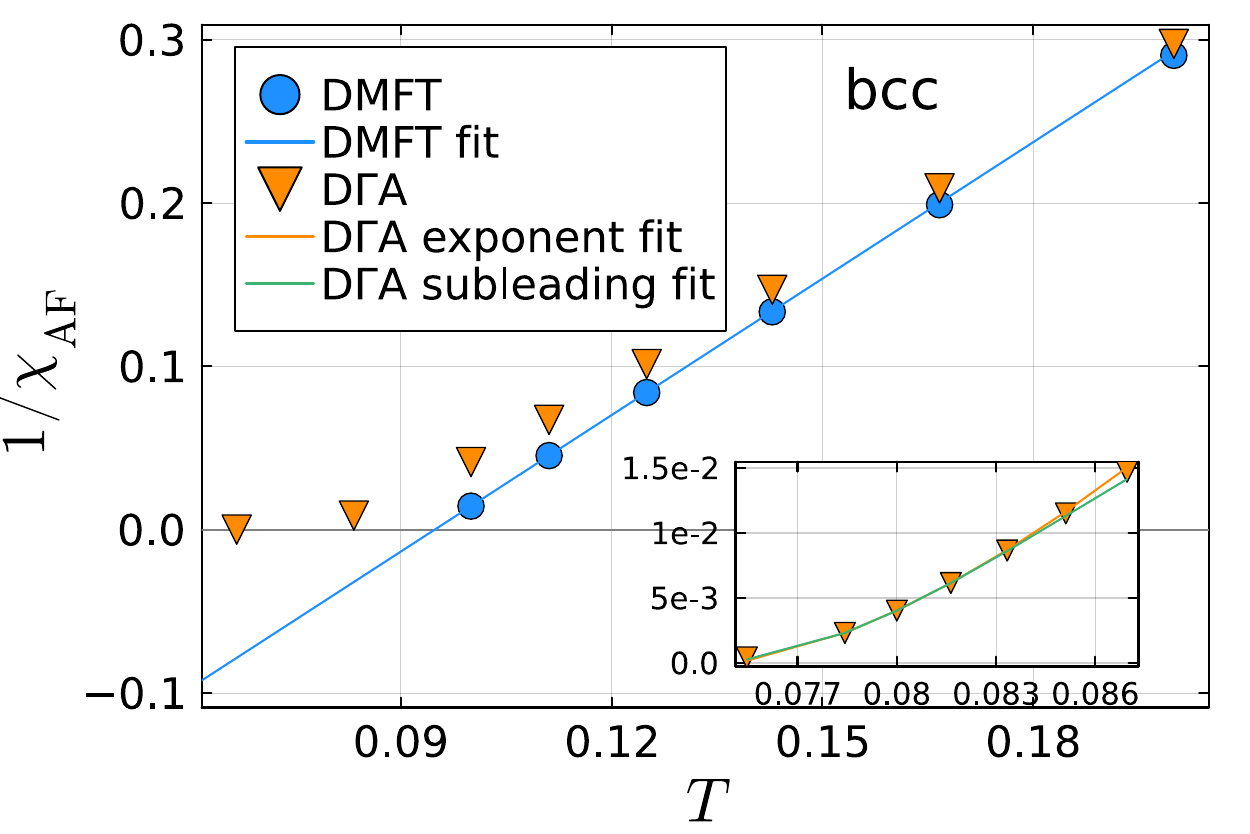}
    \includegraphics[width = 0.48\textwidth]{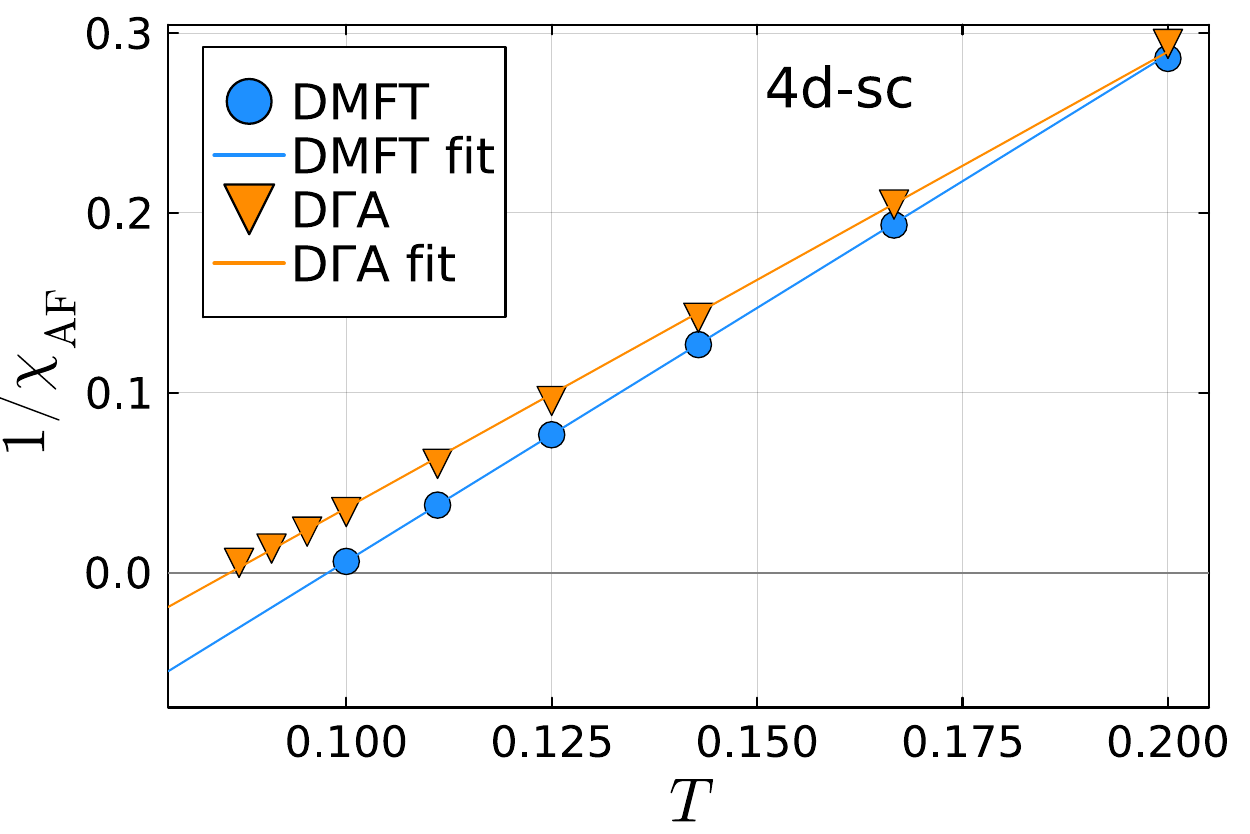}
    \includegraphics[width = 0.48\textwidth]{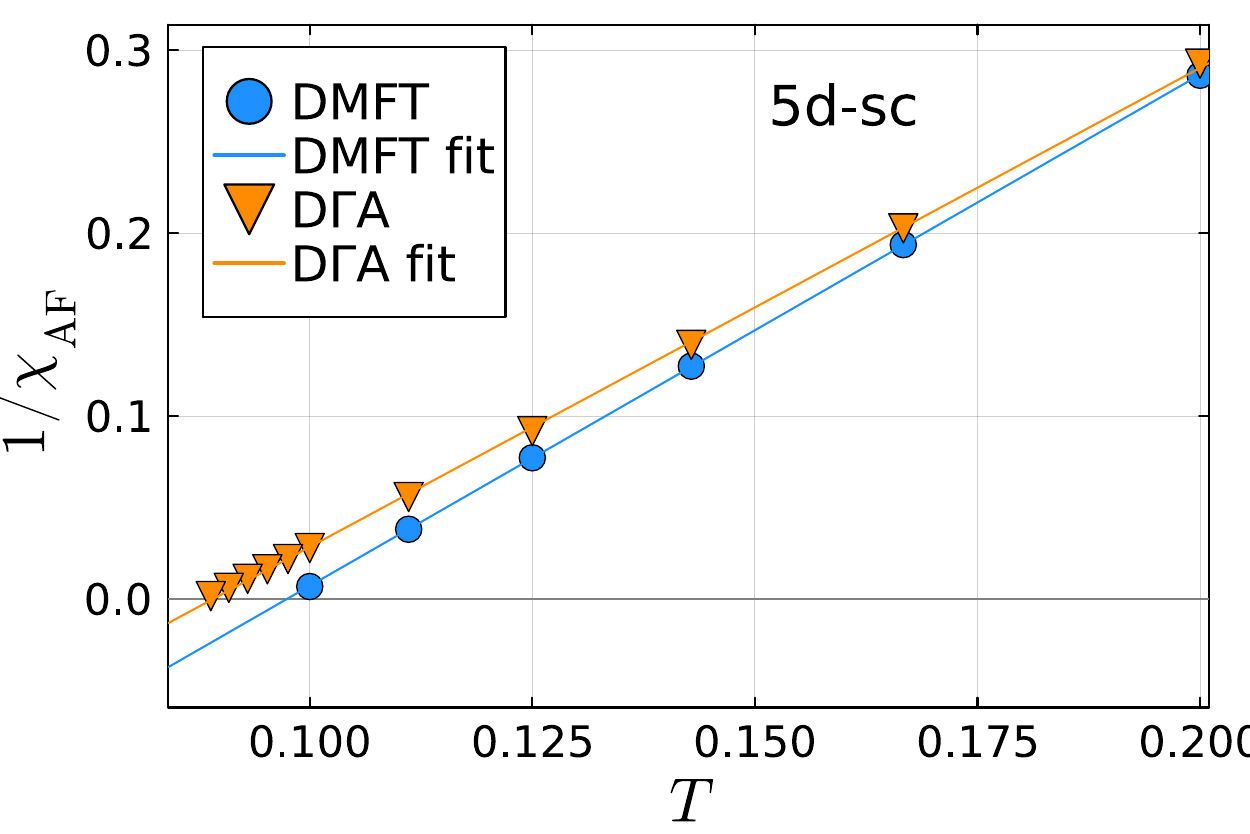}

    \caption{Inverse antiferromagnetic susceptibility $1/\chi_\text{AF}(T)=1/\chi_{m,\mathbf{q}=\mathbf{q}_N}^{\omega=0}(T)$ as function of the temperature $T$ at $U=2.0$ for the 3d-sc [$\mathbf{q}_N=(\pi,\pi,\pi)$, top left panel], bcc [$\mathbf{q}_N=(2\pi,2\pi,2\pi)$, top right panel], 4d-sc [$\mathbf{q}_N=(\pi,\pi,\pi,\pi)$, bottom left panel] and 5d-sc [$\mathbf{q}_N=(\pi,\pi,\pi,\pi,\pi)$, bottom right panel] lattice, respectively. DMFT and D$\Gamma$A results are indicated by blue circles and orange triangles. Blue lines represent linear fits to the DMFT data. For the 3d-sc and the bcc lattices, the inset shows fits to the numerical lD$\Gamma$A data for the two fitting functions in Eqs.~\eqref{equ:scalingfunction} and \eqref{equ:fitsubleading}. For the 4d-sc and 5d-sc lattices the orange line is a linear fit to the lD$\Gamma$A data. %\gr{The y label of the figure should be $1/\chi_\text{AF}$ indstead of $1/\chi_m$. Moreover, you could put the lattice type as a label into the figure which makes it easier for the reader.}
    }
    \label{fig:bipartite_inv-susc}
\end{figure*}

{\sl fcc lattice:} The results for the frequency and momentum dependence of the spin susceptibility for the fcc lattice are presented in Figs.~\ref{fig:fcc-susc} and \ref{fig:fcc-susc-q}, respectively.
%\gr{Maybe, we can show her in insets results for high temperatures to demonstrate the effect of geometric frustration. With the results for low temperatures alone this is somehow difficult to discuss.}
At high temperatures (inset in the left panel of Fig.~\ref{fig:fcc-susc}), the peak in the local spin susceptibility $\chi_{m,\text{loc}}^\omega$ at $\omega\!=\!0$ is considerably reduced compared to the bipartite lattices in the left panels of Fig.~\ref{fig:bipartite_susc}.
This indicates a strong suppression of magnetic fluctuations which is consistent with the  geometric frustration and the absence of nesting in the fcc lattice.
The peak at $\omega=0$ can be, however, increased by substantially lowering the temperature beyond the the ones considered for the 3d-sc, bcc, 4d-sc and 5d-sc in Fig.~\ref{fig:bipartite_susc}.
At $\beta=20.0$ (left panel of Fig.~\ref{fig:fcc-susc}) we observe a moderate value of the local spin susceptibility at $\omega=0$ which is roughly equivalent for DMFT and D$\Gamma$A indicating that nonlocal correlations are rather insignificant at this temperature.
On the contrary, at $\beta=150$ (right panel) we see a pronounced difference between DMFT and D$\Gamma$A albeit no phase transition is observed at such low temperature even in DMFT, in strong contrast to the bipartite lattices.
%We find a local spin susceptibility with a large values at $\omega\!=\!0$ in Fig.~\ref{fig:fcc-susc} and a pronounced peak at $\mathbf{q}\!=\!(0,0,2\pi)$ in the static spin susceptibility along a high-symmetry path in Fig.~\ref{fig:qpaths} which corresponds to a layered antiferromagnetic pattern where planes with spin up and spin down alternate. \gr{What would be really great is, to show also the spin-ordered pattern in the insets of Fig.~\ref{fig:qpaths}.}

The static spin susceptibility $\chi_{m,\mathbf{q}}^{\omega=0}$ at high temperatures (not shown), on the other hand, is rather flat as a function of $\mathbf{q}$ at n = 1.0 without any pronounced features at a specific wave vector. 
This is again a consequence of the strong geometric frustration in the lattice.
Hence, in Fig.~\ref{fig:fcc-susc-q}, we present data for $\chi_{m,\mathbf{q}}^{\omega=0}$ of the fcc lattice for $n=0.75$ where at a rather low temperature of $\beta=80.0$ this correlation function features a clear peak at $\mathbf{q}=\mathbf{q}_N=(0,0,2\pi)$ indicating a tendency to a magnetic ordering at this momentum vector.
We note that such an ordering vector corresponds to a layered antiferromagnetic pattern where planes with spin up and spin down perpendicular to the $q_z$ direction alternate.
This is consistent with predictions from RPA and salve-boson studies\cite{Igoshev2015} for the ground state at $n=0.75$ (see also appendix~\ref{sec:rpa}).

\subsection{Temperature dependence of the spin susceptibility}
\label{sec:criticalexponents}

In this section, we analyze the temperature dependence of the static ($\omega\!=\!0$) spin susceptibility at the ordering vector $\mathbf{q}_N$ (which is equivalent to the nesting vector for the bipartite case) for selected interaction strengths.
For convenience, we will denote this quantity as antiferromagnetic susceptibility $\chi_\text{AF}(T)=\chi_{m,\mathbf{q}=\mathbf{q}_N}^{\omega=0}(T)$, also for the fcc lattice where the geometric structure of the order does not correspond to a textbook G-type antiferromagnet due to geometric frustration.
Close to the transition temperature to the magnetically ordered state $\chi_\text{AF}(T)$ follows the universal scaling law:
\begin{equation}
\label{equ:scalingfunction}
\chi_\text{AF}(T)=\chi_{m,\mathbf{q}=\mathbf{q}_N}^{\omega=0}(T)\sim A(T-T_N)^{-\gamma}
\end{equation}
where $A$ is a constant, $T_N$ the transition temperature to the ordered state, and $\gamma$ the critical exponent associated with a second-order phase transition.
% Note that the label "AF" indicates that the susceptibility is evaluated for the wave vector at which the order is found which is typically the antiferromagnetic transfer momentum in the bipartitet case.
The functional form in Eq.~\eqref{equ:scalingfunction} implies that the susceptibility features a power law divergence at $T\!=\!T_N$.
We present our results for the bipartite and the fcc lattices in the two following separate paragraphs.

\begin{figure*}[t]
     \includegraphics[width=0.32\linewidth]{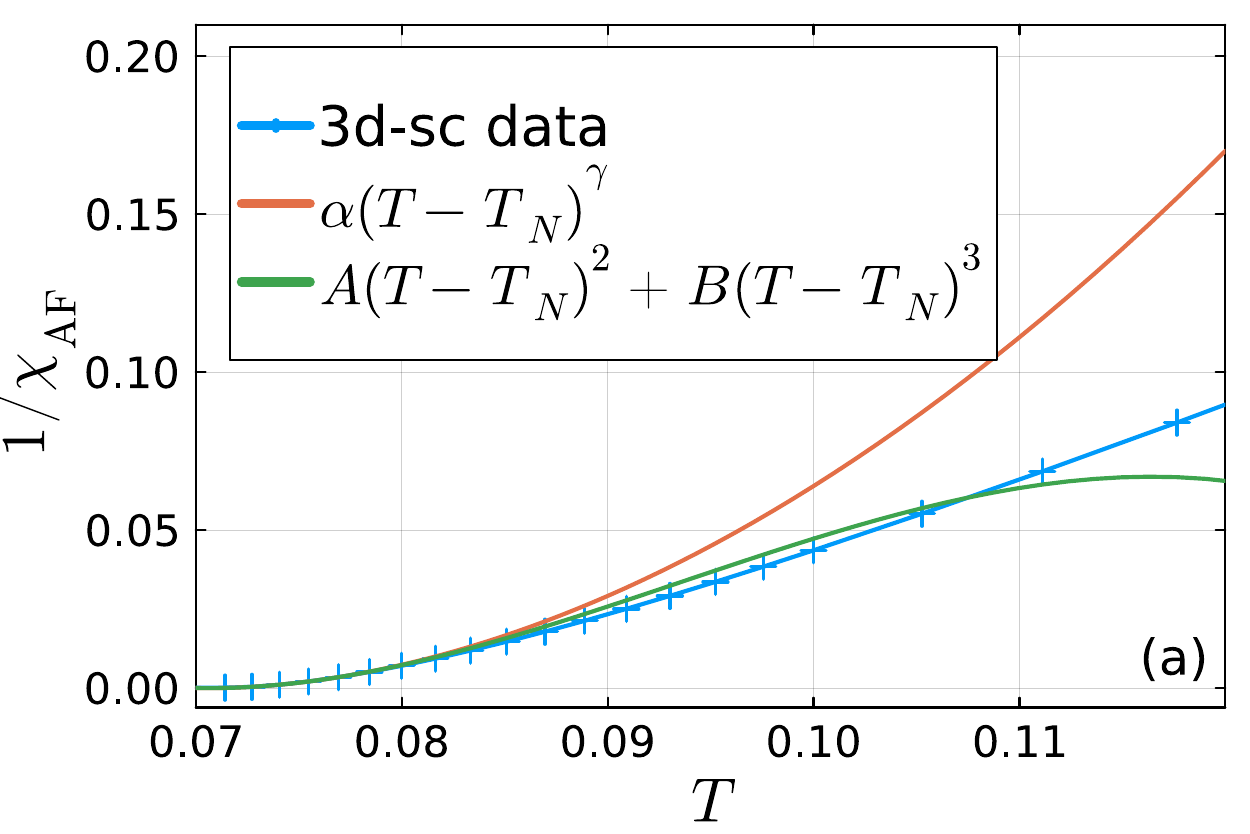} \includegraphics[width=0.32\linewidth]{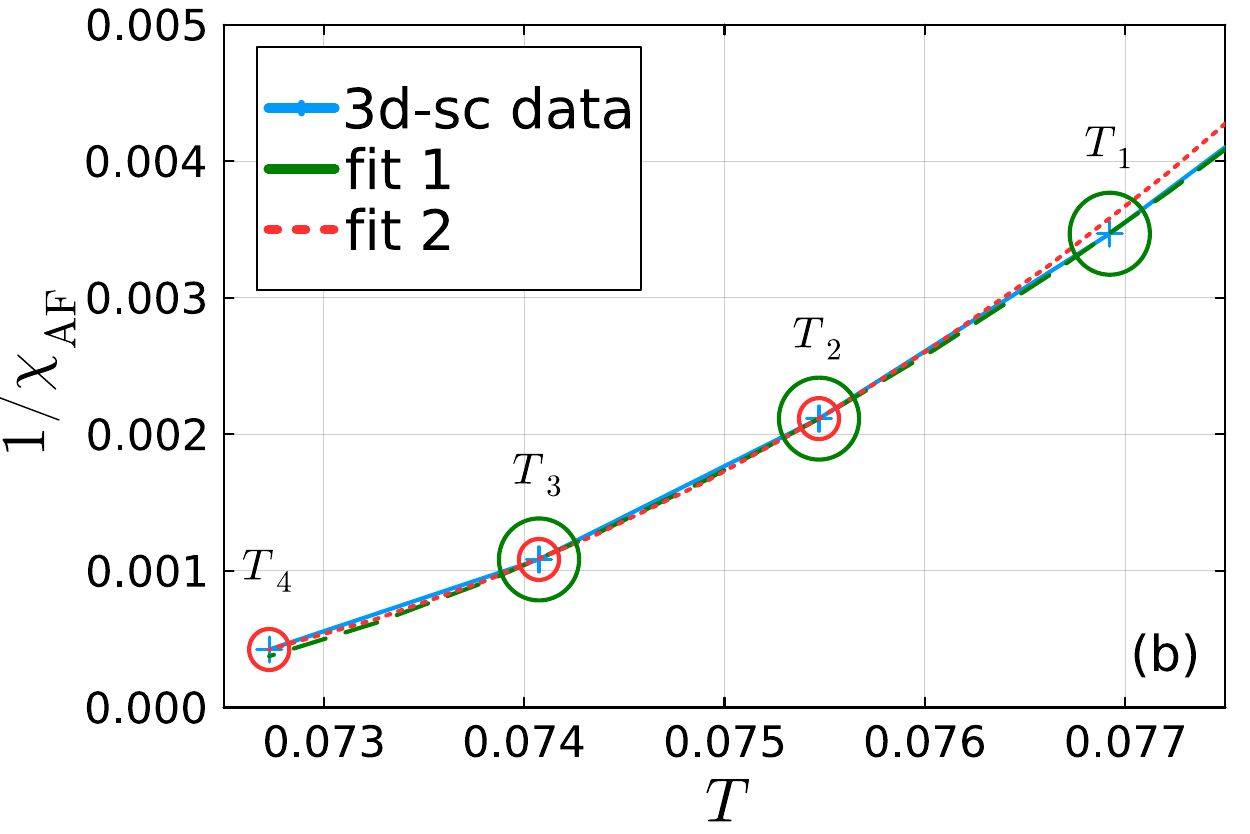}
    \includegraphics[width=0.32\linewidth]{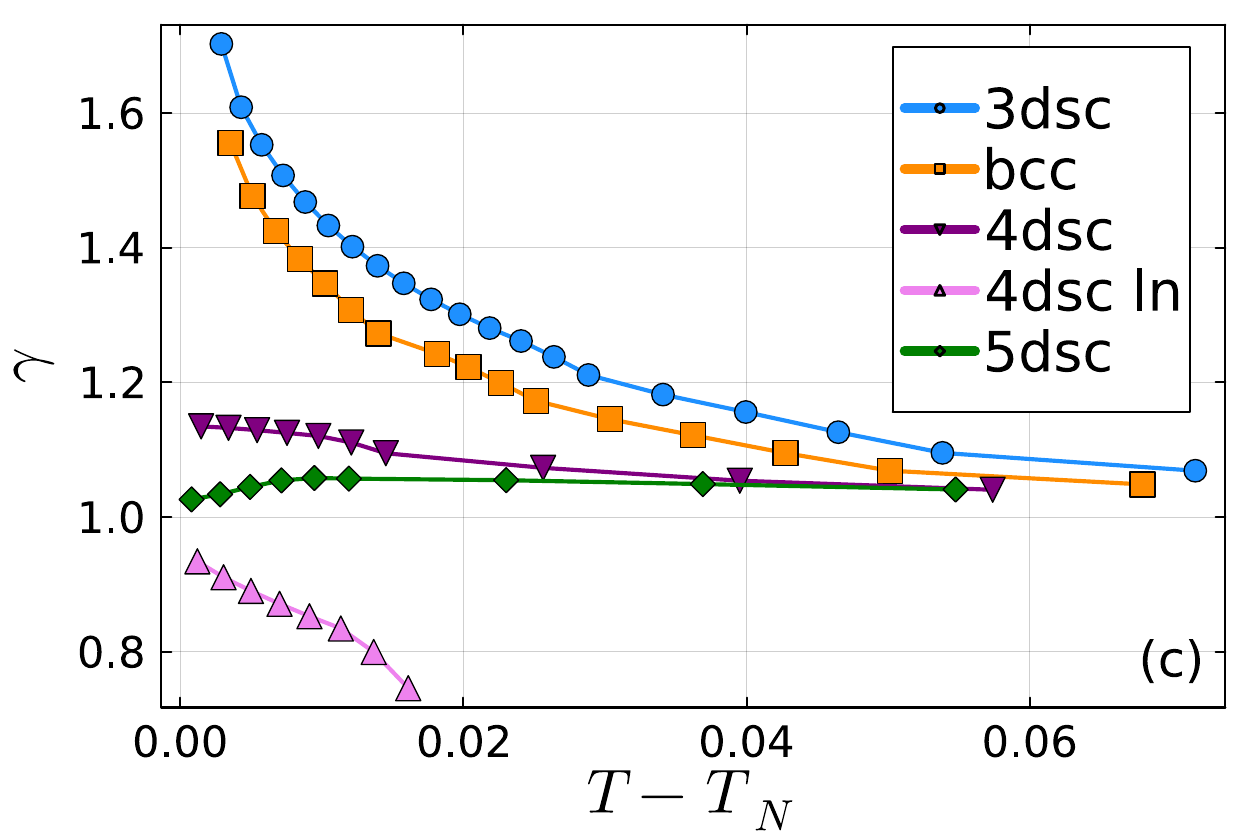}
    \caption{(a) Numerical lD$\Gamma$A data for $\chi_\text{AF}^{-1}(T)$ for the 3d-sc lattice at $U=2.0$ (blue crosses) fitted to Eqs.~\eqref{equ:scalingfunction} (orange line) and \eqref{equ:fitsubleading} (green line). (b) Illustration of the fitting procedure (for the 3d-sc lattice) used to obtain the critical exponent $\gamma$ in Eq.~\eqref{equ:scalingfunction} where only three data points are used for the fit indicated by large green and small red circles which gradually approach $T_N$. (c) Fitted value of the critical exponent $\gamma$ as a function of the distance from $T-T_N$ where the fit has been performed according to the procedure outlined in (b). Data in (b) and (c) correspond to $U=2.0$. For the 4d-sc we present data from fits without (violet lower triangles) and with (pink upper triangles) logarithmic corrections. %\gr{When writing the text to explain the procedure it seems to be easier with the following modification of the middle panel: Remove the highest data point and consider only 4 points. Then we have only violet and green circles. Moreover, put labels $T_1$ (for the highest temperature), $T_2$ and so on to the data points (see also the description in the main text). Finally, you could also cut the temperatures scale for the leftmost panel at $T_N$ as lower values are not physical.}
    %On the left, we see an example fit of the inverse susceptibility of the 3d-sc lattice at $U=2.0$. Two different fit functions are used, one fitting the critical exponent in orange and one fitting the leading and sub-leading term (l+sl) of the Taylor expansion in green. In the center we illustrate the method used to obtain the scaling of the critical exponent, when $T \rightarrow T_N$. Each fit uses three data points, circled in the same color as the resulting fit function. This example is again, for the 3d-sc lattice at $U=2.0$. On the right, we present the result of this method for the 3d-sc, bcc and 4d-sc lattice at $U=2.0$. $T$ always corresponds to the lowest temperature used for each fit.}
    }
    \label{fig:crit_exp_scaling}
\end{figure*}

%\gr{We should give data here for the bipartite case, not only Fig.~7. Moreover, for comparison the DMFT data should be included (one could make three panels for 3d-sc, 4d-sc and bcc which also includes the DMFT results instead of Fig.~7).}

{\sl Bipartite lattices:} In Fig.~\ref{fig:bipartite_inv-susc} we show our data for the inverse antiferromagnetic spin susceptibility $\chi_\text{AF}^{-1}(T)$ as function of temperature $T$ for the bipartite 3d-sc, bcc, 4d-sc and 5d-sc lattices.
Blue circles correspond to DMFT results, while our lD$\Gamma$A data are represented by orange triangles.
The temperature at which $\chi_\text{AF}^{-1}(T)$ crosses zero corresponds to the transition temperature $T_N$ to the magnetically ordered state.
Within DMFT $\chi^{-1}_\text{AF}(T)$ is a straight line for all bipartite lattices which is consistent with the standard mean-field critical exponent $\gamma\!=\!1$.
The nonlocal correlations included by D$\Gamma$A affect the antiferromagnetic susceptibility in two ways:
(i) As expected, at high temperatures our lD$\Gamma$A data almost coincide with the DMFT results for $\chi_\text{AF}^{-1}(T)$.
%\sout{This is not surprising as DMFT becomes exact in the limit of infinite temperature where nonlocal correlations induced by lD$\Gamma$A are negligible.}
As $T$ decreases the lD$\Gamma$A data for $\chi_\text{AF}^{-1}(T)$ start to deviate from the DMFT results while still keeping a linear dependence on $T$ albeit with a reduced slope. 
Already this deviation from DMFT leads to a zero-crossing of the lD$\Gamma$A $\chi_\text{AF}^{-1}(T)$ at lower temperatures than in DMFT, hence, featuring a reduced critical temperature.
For the 3d-sc and the bcc lattices (upper left and right panel, respectively) this reduction of $T_N$ gets further enhanced by (ii) a deviation from the linear shape at low temperature close to the phase transition, where the critical exponent $\gamma$ becomes  visually larger than one.
The temperature region $\Delta T_\text{crit}\!=\!T-T_N$ where such a deviation from the linear behavior can be observed can be estimated via the Ginzburg-Landau criterion\cite{Landau1937original,*Landau1937Collection} which yields $\Delta T_\text{crit}\propto\!T_N^2$.
Let us note, however, that the critical region is smaller for the bcc with respect to the 3d-sc lattice which is consistent with the fact that the bcc lattice has a larger coordination number and, hence, DMFT may be expected to be more accurate for this system in a broader parameter regime.
%\gr{Maybe we show current Fig.~\ref{fig:critical_exponent_4d-sc} as an inset of the middle panel of Fig.~\ref{fig:bipartite_inv-susc} as it does not contain a lot of new information. The question is then only whether we should consistently show data for $U\!=\!2.0$ (in Fig.~\ref{fig:critical_exponent_4d-sc} we have $U\!=\!1.0$).}
For the 4d-sc in the lower left panel of Fig.~\ref{fig:bipartite_inv-susc} our numerical lD$\Gamma$A data indicate that $\gamma=1$ coinciding with the mean-field prediction.
This is indeed the expected behavior in four dimensions although logarithmic corrections can be expected as $d=4$ corresponds to the upper critical dimensions.
These logarithmic corrections can make the numerical fitting procedure more challenging which we will discuss below in more detail.
%Here, only logarithmic corrections to the mean-field exponents can be expected (which cannot be resolved within our numerical precision). 
In any case, the sizable reduction of $T_N$ observed for the 4d-sc lattice originates exclusively from feature (i), i.e., the rigid shift of $\chi_\text{AF}^{-1}(T)$ by lD$\Gamma$A discussed above. 
%In the absence of logarithmic corrections, for the 5d-sc lattice, the linearity of the lD$\Gamma$A results is even more pronounced.
Finally, for the 5d-sc lattice a deviation from the linear temperature dependence of $\chi_\text{AF}^{-1}(T)$ predicted by DMFT is {\sl not} observed in our lD$\Gamma$A data (lower right panel of Fig.~\ref{fig:bipartite_inv-susc}), thus yielding $\gamma\sim 1$.
%\sout{confirms the general understanding} 
This is fully consistent with the general understanding that for $d\ge5$ mean-field critical behavior is recovered as the exact solution of the problem.

In a next step, we have analyzed the temperature dependence of $\chi_\text{AF}^{-1}(T)$ close to $T_N$ more accurately for the 3d-sc and the bcc lattices, where evident deviations from the mean-field exponent are observed. 
This has allowed us to investigate the critical behavior more thoroughly and, in particular, to provide accurate estimates for the critical temperature $T_N$ and the critical exponent $\gamma$ in Eq.~\eqref{equ:scalingfunction}.
%\gr{Maybe we can also show data for the 3d-sc lattice here?}
To this end, we have fitted our numerical data for $\chi_\text{AF}^{-1}(T)$ to Eq.~\eqref{equ:scalingfunction}.  
The results of these fits, which are shown as orange lines in the insets of the upper panels panels of Fig.~\ref{fig:bipartite_inv-susc} for the 3d-sc and bcc lattice, respectively, as well as for the 3d-sc on a larger scale in the left panel (a) of Fig.~\ref{fig:crit_exp_scaling},
predict a critical exponent of $\gamma=1.556$ for the bcc and $\gamma\!=\!1.7026$ for the 3d-sc lattice.
The latter value for the 3d-sc lattice is in good agreement with previous ladder D$\Gamma$A\cite{Rohringer2011} and dual fermion\cite{Hirschmeier2015} calculations for this lattice.

Let us, however, note that the actual critical exponent can be precisely observed only extremely close to the phase transition, while at larger values of $T$ deviations from the scaling behavior in Eq.~\eqref{equ:scalingfunction} emerge.
Fitting our numerical results, which due to the accessible range partly include such data at higher temperatures, to this scaling function will then result in an inaccurate determination of the critical exponent $\gamma$.
To mitigate this problem and obtain more reliable results we have, hence, devised the following procedure illustrated in Fig.~\ref{fig:crit_exp_scaling}(b), which allows us to gradually improve the accuracy of the numerical value for $\gamma$:
We start at rather high temperatures far away from $T_N$ by fitting the numerical data for $\chi_\text{AF}^{-1}(T)$ to Eq.~\eqref{equ:scalingfunction} using only three data points corresponding to the temperatures $T_1$, $T_2$ and $T_3$ which are indicated by green circles in Fig.~\ref{fig:crit_exp_scaling}(b).
We denote the obtained value of $\gamma$ as $\gamma(T_3)$ since $T_3$ is the last point of this data set. 
%\ml{We use the last point of the data set, I believe this makes more sense for the $T -T_N$ scale.}
In the next step, we replace the data point at the hightest temperature $T_{1}$ with the data point at $T_4$ which is one step lower than the lowest previous data point at $T_3$.
Hence, the new data set which we use for the next fit consists of the values for $\chi_\text{AF}^{-1}(T)$ at the temperatures $T_2$, $T_3$ and $T_4$ [indicated by red circles in Fig.~\ref{fig:crit_exp_scaling}(b)] and gives us a new critical exponent $\gamma(T_4)$.
Iterating this procedure we obtain a function $\gamma(T=T_i)$ where $T_i$ always corresponds to the lowest of the three temperatures.
We have plotted this function $\gamma(T)$ relative to $T_N$ in Fig.~\ref{fig:crit_exp_scaling})(c) for the 3d-sc (blue circles), bcc (orange squares), 4d-sc (violet lower and pink upper triangles) and 5d-sc (green diamonds) lattices.
% However, as the real critical exponent can be observed only at the phase transition, we have carried out a series of fits where the largest temperature $T_\text{max}$ which we consider for the fit is gradually reduced. 
% Figure~\ref{fig:crit_exp_scaling} shows the critical exponent $\gamma$ as a function of $T_\text{max}$ for the 3d-sc- and bcc lattice.
For the 3d-sc and bcc lattices we clearly see that $\gamma(T)$ has not saturated at the smallest temperatures but, on the contrary, is very steep which indicates that a larger critical exponent $\gamma$ than the previously fitted values of $\gamma=1.7026$ for the 3d-sc and $\gamma=1.556$ for the bcc lattice can be expected.
%the curve is very steep which indicates that a larger critical exponent than $\gamma\!=\!1.7026$ can be expected for the sc lattice ($\gamma\!=\!1.556$ for the bcc lattice). 
In fact, a fit of the curve $\gamma(T)$ to $T\!\rightarrow\!T_N$ predicts values of $\gamma\!\sim\!1.89$ for the 3d-sc lattice and $\gamma\!\sim\!1.76$ for the bcc lattice.
It seems likely that these values might further increase if we could evaluate $\chi_\text{AF}^{-1}(T)$ for temperatures even closer to $T_N$ (which is numerically infeasible).

The above considerations have motivated us to consider the question of the critical exponent $\gamma$ for the 3d-sc and the bcc lattice obtained within lD$\Gamma$A from a more general perspective. 
Indeed, it was numerically observed in Ref.~\cite{DelRe2019}, where lD$\Gamma$A calculations were performed for the attractive Hubbard model on a 3d-sc lattice, that the fitted exponents appeared compatible with the universality class of the Berlin-Kac model where the number of components of the order parameter $N\!\rightarrow\!\infty$ (corresponding to a mean field in the dimension of the order parameter), similar as those of the two-particle-self-consistent approach \cite{Dare1996}.
%\sout{The method of $\lambda$ corrections becomes exact in the limit of an infinite dimensional order parameter\cite{Dare1996} known as the Berlin-Kac model.
%In this case, critical exponents can be calculated analytically and give $\gamma\!=\!2$.}
In fact, the value of $\gamma\!=\!2$ associated to the Berlin-Kac universality class~\cite{Dare1996} appears well compatible also with our numerical data, when we fix $\gamma=2$ and include a subleading term proportional to $(T-T_N)^3$ in our numerical fit function
\begin{equation}
\label{equ:fitsubleading}
\chi_\text{AF}^{-1}(T)=A(T-T_N)^2+B(T-T_N)^3.
\end{equation}
We have tested this fit function for the 3d-sc and the bcc lattice in the insets of the upper panels of Fig.~\ref{fig:bipartite_inv-susc} and Fig.~\ref{fig:crit_exp_scaling}(a) (green curve), finding an excellent agreement with our numerical data.
This indicates that, in three dimensions,  lD$\Gamma$A calculations with infinite numerical precision (not feasible in practice) might indeed yield the value of $\gamma=2$ of the Berlin-Kac model, thus overestimating the expected exact value for the 3d-Heisenberg universality class ($\gamma \simeq 1.41$).

For the 4d-sc, we expect a critical exponent $\gamma=1$ (with logarithmic corrections) as we are at the upper critical dimension.
Our numerical data for $\gamma(T)$ for the 4d-sc lattice in Fig.~\ref{fig:crit_exp_scaling}(c) indeed shows only a weak dependence of $\gamma(T)$ on $T$, which varies between $1.0$ and $1.15$.
To better understand the small deviation from the expected value of $\gamma\!=\!1$ we have made a second fit using a fit function which includes the logarithmic corrections\cite{Bauerschmidt2014}
\begin{equation}
\label{equ:fitwithlogarithmic}
\chi_\text{AF}(T)=\chi_{m,\mathbf{q}=\mathbf{q}_N}^{\omega=0}(T)\sim A(T-T_N)^{-\gamma}\ln\left(\frac{1}{T-T_N}\right)^{\hat{\gamma}}
\end{equation}
where we have set $\hat{\gamma}=1$ which corresponds to the value expected for the Berlin-Kac model.
With this ansatz our values for $\gamma$ become smaller than $1$ [upper pink triangles in Fig.\ref{fig:crit_exp_scaling}(c)] and rapidly approach the expected value of $1$ from below.
Hence, although the fitting procedure is numerically delicate in $4d$ we conclude that our numerical data are compatible with the expected value $\gamma=1$ when logarithmic corrections are included in the fit function.

% \gr{In this respect let us stress, that from an analytical perspective the critical exponent $\gamma$ of $\lambda$-lD$\Gamma$A has to take a value of $\gamma=1$ in $4d$ (apart from logarithmic corrections).} Taking these into account still gives a small deviation from the linear critical exponent ($\gamma \approx 0.93$). However, in this case we can see the critical exponent tending closer to $1$ for $T\rightarrow T_N$.
% \gr{While we were not able to unambiguously identify the source of the deviation of our numerical data from this result we speculate that it originates from the limited frequency grid for both the determination of the two-particle vertex function and the evaluation of the sum rule in Eq.~\eqref{eq:pauli_principle} and/or the finite number of bath sites in our exact diagonalization solver which both cannot be increased due to numerical limitations.}
%These small deviations from the expected $\gamma=1.0$ can be attributed to finite numerical precision when evaluating momentum integrals in a four dimensional momentum space.
Finally, for the 5d-sc lattice [green diamonds in Fig.~\ref{fig:crit_exp_scaling}(c)] we observe a very good agreement with the expected mean-field critical exponent $\gamma=1$ close to the phase transition. 
The very small deviations can be attributed to the finite numerical precision and the numerical limitations for the size of the momentum grids in $5d$.

% where no logarithmic corrections occur, shows very small deviations from $\gamma = 1$, only because we're numerically unable to reach $T_N$ exactly. This is the expected behavior.

\begin{figure}[t!]
    \centering
    \includegraphics[width=0.48\linewidth]{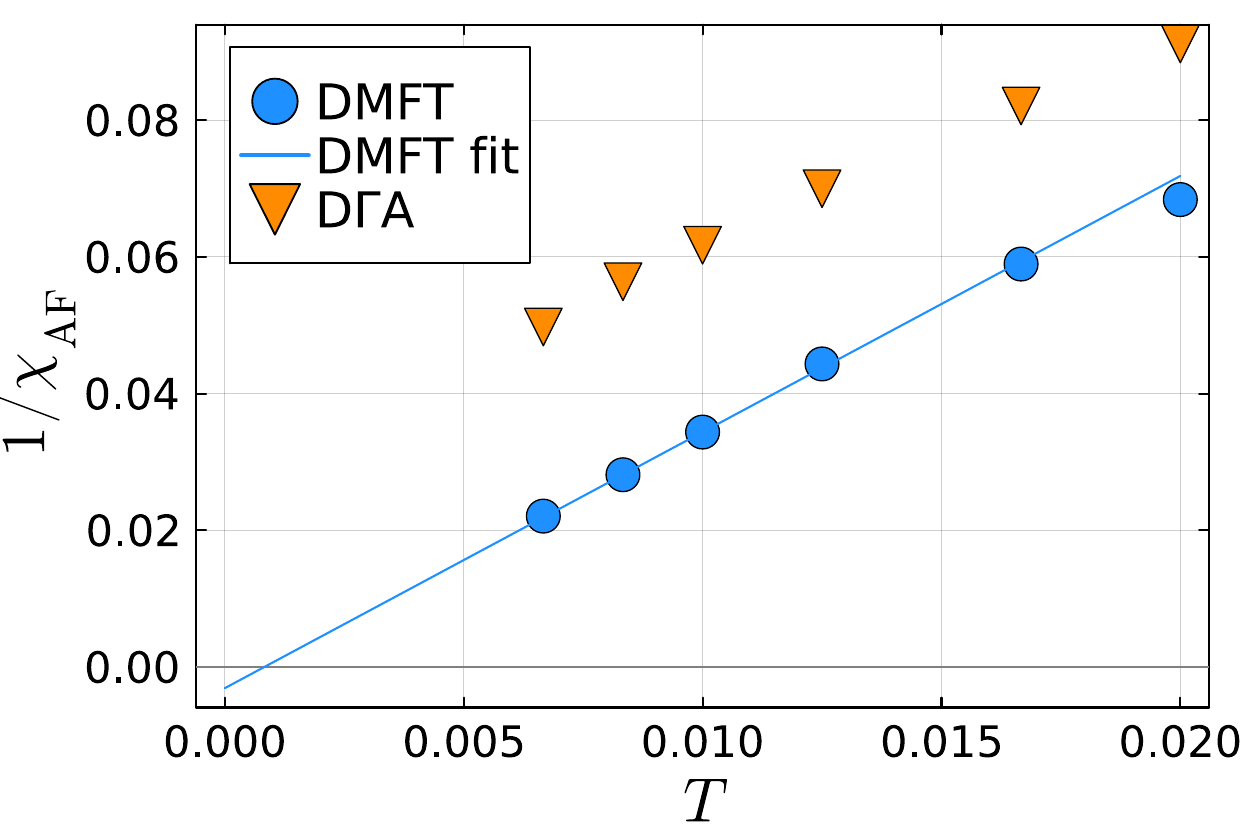}
    \includegraphics[width=0.48\linewidth]{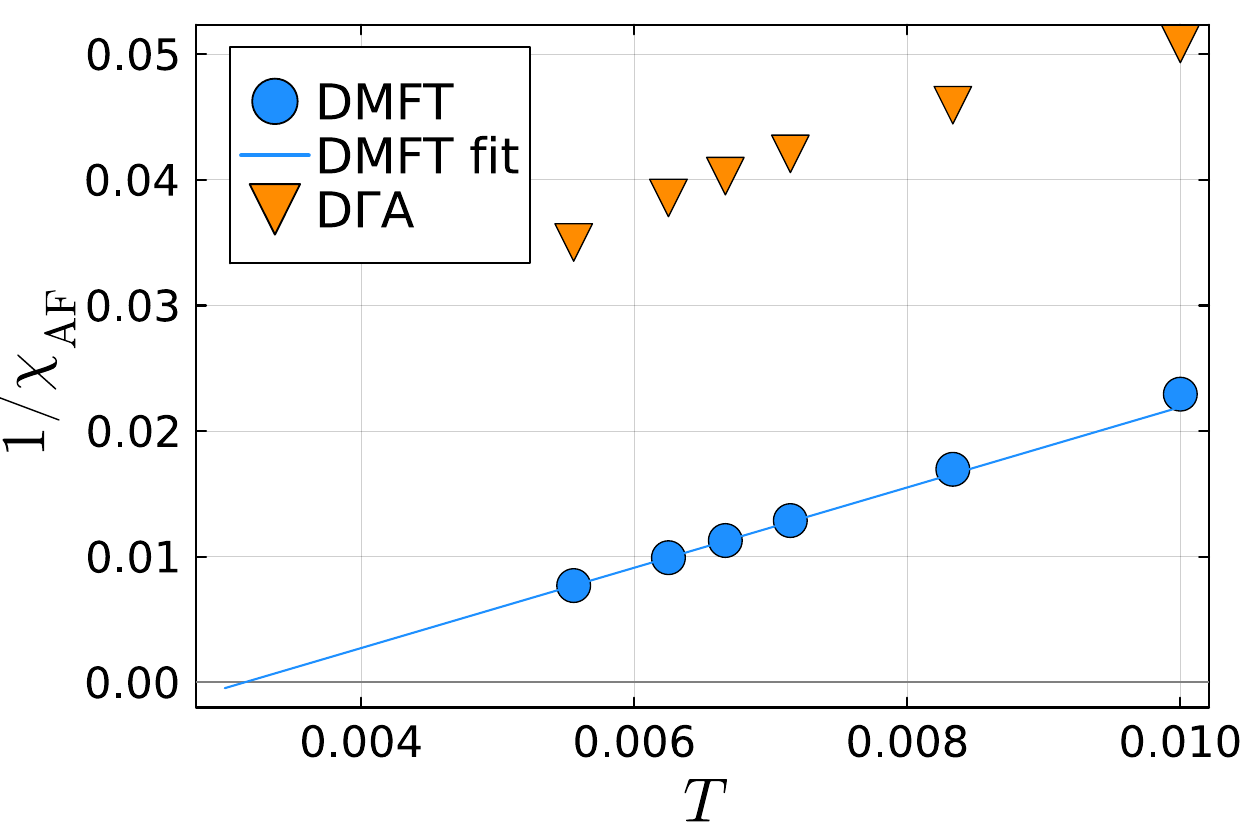}
    \caption{Inverse antiferromagnetic susceptibility $1/\chi_\text{AF}(T)=1/\chi_{m,\mathbf{q}=\mathbf{q}_N}^{\omega=0}(T)$ for $\mathbf{q}_N=(0,0,2\pi)$ as a function of $T$ for the fcc lattice at $U=3.0$ (left panel) and $U=5.0$ (right panel) for $n=0.75$. The blue line is a linear fit to the DMFT data.}
    \label{fig:inv_susc_fcc}
\end{figure}

{\sl fcc lattice:} The results for $\chi_\text{AF}^{-1}(T)$ in the fcc lattice are depicted in Fig.~\ref{fig:inv_susc_fcc} for the two different interaction strengths $U=3.0$ (left panel) and $U=5.0$ (right panel) at $n\!=\!0.75$. 
As already discussed in the previous section, spin fluctuations are strongly suppressed in this lattice due to the large geometric frustration.
In order to observe an appreciable antiferromagnetic susceptibility, we have to consider very low temperatures and large interaction strengths where numerical calculations are notoriously difficult.
Within DMFT (blue circles) we find the expected linear behavior of $\chi_\text{AF}^{-1}(T)$ for both considered interaction strengths.
%The crossings of these lines with zero indicates the DMFT transition temperature.
We can see that for $U=3.0$ in the left panel $T_N\simeq 0$ which indicates that no magnetic order can be observed in the fcc lattice for $U\lesssim 3.0$.
Hence, while in the bipartite lattices a magnetic order is observed for all interaction strengths both in DMFT and D$\Gamma$A such an order emerges in the fcc lattice only above a critical interaction strength $U_c\!~\approx\!3.0$ in DMFT.
%Let us, however, not that an unambiguous identification of the ordered phase is numerically difficult due to the extremely small values of $T_N$.

Remarkably, the inclusion of nonlocal correlations obtained via lD$\Gamma$A (orange triangles) {\sl completely} suppresses the antiferromagnetic order predicted by DMFT,  as no zero-crossing of $\chi_\text{AF}^{-1}(T)$ can be anticipated from the extrapolation of this correlation function to small temperatures.
It appears that the D$\Gamma$A results are also linear in $T$, which would be consistent with the large coordination number of 12 for the fcc lattice where a mean-field exponent can be expected in a broad region of the phase diagram.
Let us, however, mention that the determination of $\lambda$ at very low temperatures is numerically extremely challenging, which limits the conclusions we can draw on the $T \rightarrow 0$ magnetic properties in the fcc lattice.
%\sout{which prevents definite conclusions about the D$\Gamma$A antiferromagnetic susceptibilities for the fcc lattice.}

\subsection{Phase diagram}
\label{sec:phasediagram}

\begin{figure}[t]
    \centering
    \includegraphics[width=1.0\linewidth]{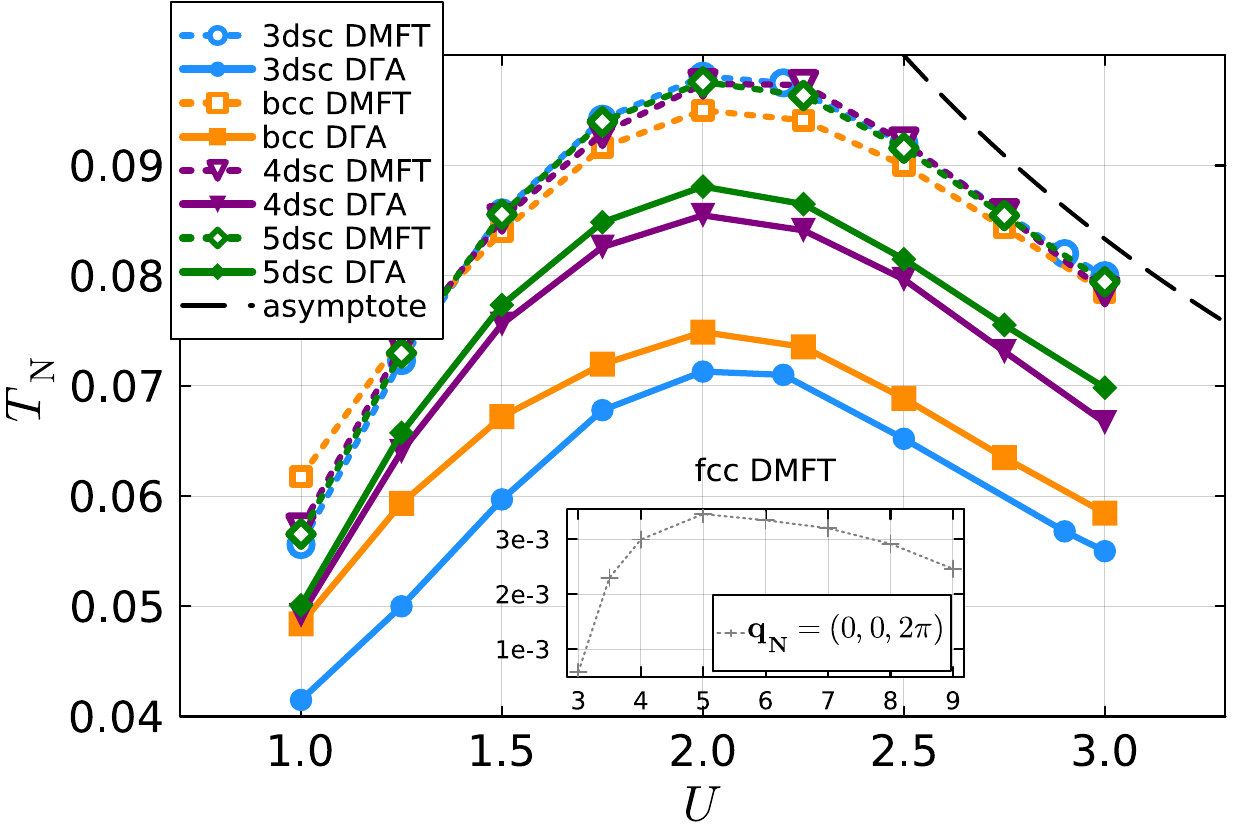}
    \caption{%\gs{I wonder if we should add some kind of "key" or mini-title for the inset. As it is now, it looks very "out-of-context".} 
    Phase diagram for the 3d-sc (blue circles), bcc (orange squares), 4d-sc (violet triangles) and 5d-sc (green diamonds) lattices obtained from DMFT (empty symbols/dashed lines) and lD$\Gamma$A (filled symbols/solid lines). The inset shows the (approximate) $T_N(U)$ obtained by DMFT for the fcc lattice.}
    \label{fig:phase_diagram}
\end{figure}

\begin{table}
\renewcommand{\tabcolsep}{4pt} 
\begin{tabularx}{\textwidth}{c |c c c c c c c c c } 
$U$ & 1.0 & 1.25 & 1.5 & 1.75 & 2.0 & 2.25 & 2.5 & 2.75 & 3.0 \\ 
\hline
3d-sc & 25.4 & 30.8 & 30.3 & 28.0 & 27.3 & 27.2 & 29.1 & 30.7 & 31.2 \\
bcc & 21.6 & 20.2 &  20.0 &  21.5 & 21.2 &  21.9 & 23.5 & 24.7 & 25.5 \\
4d-sc & 13.8 & 13.0 & 11.2 & 11.1 & 12.2 & 13.6 & 13.5 & 14.6 & 15.1 \\
5d-sc & 11.4 & 9.9 & 9.6 & 9.7 & 9.7 & 10.2 & 11.0 & 11.6 & 12.1
\end{tabularx}
\caption{ Reduction of the DMFT N\'eel temperature by lD$\Gamma$A (in percent) at different values of $U$ for the four bipartite lattices under consideration.}
\label{tab:percentage_reduction}
\end{table}

In this section, we determine the transition curve $T_N(U)$ by extracting the values of $T_N$ at which $\chi_\text{AF}^{-1}(T)$ crosses zero for a range of interaction values $U$.
%By determining the temperature $T_N$ at which $\chi_\text{AF}^{-1}(T)$ crosses zero for a range of interaction values $U$ allows us to determine the $T$ vs. $U$ phase diagrams for all lattices considered in this paper.
The results are reported in Fig.~\ref{fig:phase_diagram} where solid lines represent the D$\Gamma$A data and dashed lines correspond to the DMFT results.
In the main panel the lD$\Gamma$A N\'eel temperatures for the bipartite lattices are shown, while the fcc transition temperature is presented in the inset (for DMFT only).
            
{\sl Bipartite lattices:} For the four bipartite lattices we observe a sizable reduction of $T_N$ of DMFT (empty symbols/dashed lines) by nonlocal correlations computed in lD$\Gamma$A (filled symbols/solid lines).
The size of the reduction depends on the dimension and the coordination number of the lattice.
In fact, while we have chosen our system parameters in such a way that the transition temperatures in DMFT are (almost) equivalent for all bipartite lattices, the associated lD$\Gamma$A curves are visibly different.
The largest deviation for $T_N$ between DMFT and lD$\Gamma$A is observed for the 3d-sc lattice (blue empty and filled circles) which has a coordination number of $6$.
For the bcc lattice (orange empty and filled squares) with a coordination number of $8$, this difference is already smaller with respect to the 3d-sc lattice case, consistent with higher accuracy of DMFT for larger coordination numbers. 
%\sout{the fact that DMFT becomes more accurate for increasing coordination number.}
An even smaller difference for $T_N$ between DMFT and D$\Gamma$A is observed for the tesseract lattice, which has the same coordination number 8 as the bcc one but lives in four spatial dimensions. 
Here, the mean-field critical exponent $\gamma\!=\!1$  prevents a further reduction of $T_N$ in the critical region compared to the bcc case for which $\gamma\! >\! 1$. 
As expected, the 5d-sc lattice exhibits the smallest deviation between DMFT and lD$\Gamma$A since it has the highest coordination number ($Z=10$). 
Let us point out, that while the gap between the lD$\Gamma$A phase transition for the 3d-sc and 4d-sc is rather large the difference in $T_N$ between the 4d-sc and 5d-sc lattice is substantially smaller.
This might originate from the difference in the critical exponent $\gamma$ between three and four dimensions (while they are equivalent for four and five dimensions) and indicates a rather slow convergence to the $d\!=\!\infty$ limit where $T_N$ of lD$\Gamma$A should become equivalent to the corresponding DMFT value.
% However, the gap between the lD$\Gamma$A phase transitions of the 4d-sc and 5d-sc lattices is very small, in contrast to the gap between the 3d-sc and the 4d-sc lattices. This is most likely a result of the linear critical exponent in four and five dimensions, thus increasing $T_N$ significantly. Additionally, the small gap between the 4d-sc and 5d-sc results suggests either a very slow convergence of $T_N^{\mathrm{D\Gamma A}}\rightarrow T_N^{\mathrm{DMFT}}$ for $d \rightarrow \infty$, or more worryingly no convergence at all. This is subject to further research.
%\sout{$\gamma\!> \!2$.}  \sout{(see Sec.~\ref{sec:susclength}) leads to a reduced $T_N$.}
To better quantify our results, we provide in Table~\ref{tab:percentage_reduction} the relative reduction 
%\gs{(watch out how to call this -- see my comment in the caption to the table)} 
of $T_N$ between DMFT and lD$\Gamma$A for all lattice types.

Let us also analyze the asymptotic behavior of $T_N$ for large values of the interaction strength $U$. 
In this limit the (bipartite) Hubbard model can be mapped onto an Heisenberg model $H_\text{Heis}\!=\!J\sum_{\langle ij\rangle}\mathbf{S}_i\mathbf{S}_j$ via a Schrieffer-Wolff transform\cite{Schrieffer1966} where $J\!=\!\frac{4t^2}{U}>0$ 
%\gs{(most likely it is so but let's just make sure that 4 in the expression of $J$ does not depend on the dimension)} 
couples two neighboring spins antiferromagnetically.
Within DMFT this leads to an Ising type result for $T_N$ (for spin $1/2$ particles) which is given by $T_N=\frac{zt^2}{U}$ where $z$ is the coordination number of the lattice.
%Within DMFT \at{of antiferromagnetism}, the transverse fluctuations $S_i^xS_j^x\!+\!S_i^yS_j^y$ in $\mathbf{S}_i\mathbf{S}_j$ are neglected and only the $S_i^zS_j^z$ term is retained.
%\at{{\bf Vorsicht: dieser Satz hat nur Sinn in der AF-Phase!!}}
%Therefore the mean-field treatment of antiferromagnetism in DMFT becomes equivalent to the standard Weiss mean field approach for the (spin $\frac{1}{2}$) Ising model. 
%\at{{\bf Vorsicht: das gilt nur f\"ur statische Eigenschaften!!}}
%.
%In this case, the condition for a phase transition at $T_N$ is given by $\frac{zt^2}{UT_N}\!=\!1$ where $z$ is the coordination number of the lattice. 
Since in this work we have fixed $zt^2\!=\!\frac{1}{4}$ for all lattices the asymptotic behavior of the DMFT transition temperatures coincide between all bipartite lattices taking a value of $T_N\!\sim\!\frac{1}{4U}$. %\gr{Maybe you can add this curve (as black dashed line) in the figure.}
On the contrary, D$\Gamma$A takes into account transverse spin fluctuations and, hence, approaches the exact Heisenberg solution for all (bipartite) lattices which are obviously different for the different lattice types.

%\begin{table}
%\renewcommand{\tabcolsep}{4pt} 
%\begin{tabularx}{\textwidth}{c |c c c c c c c c c } 
%$U$ & 1.0 & 1.25 & 1.5 & 1.75 & 2.0 & 2.25 & 2.5 & 2.75 & 3.0 \\ 
%\hline
%3d-sc & 74.6 & 69.2 & 69.7 & 72.0 & 72.7 & 72.8 & 70.9 & 69.3 & 68.8 \\
%bcc & 78.4 & 79.8 &  80.0 &  78.5 & 78.8 &  78.1 & 76.5 & 75.3 & 74.5 \\
%4d-sc & 86.2 & 87.0 & 88.8 & 88.9 & 87.8 & 86.4 & 86.5 & %85.4 & 84.9 
%\end{tabularx}
%\label{tab:percentage_reduction}
%\caption{\gs{looking at the table, we should make clear that the first like is $U$....} Suppression \gs{(If this is the "reduction", I interpret 86.2 \% as a huge reduction, but you mean actually 13.8 \% reduction....)\ml{changed the wording to suppression, does this make sense?}} of the N\'eel temperature (in percent) at different values of $U$ for the three bipartite lattices under consideration. \at{{\bf Vorsicht: $T_N$ f\"ur 3d-sc scheint leicht inkorrekt zu sein}}}
%\end{table}
% \begin{table}
% \renewcommand{\tabcolsep}{4pt} 
% \begin{tabularx}{\textwidth}{c |c c c c c c c c c } 
% $U$ & 1.0 & 1.25 & 1.5 & 1.75 & 2.0 & 2.25 & 2.5 & 2.75 & 3.0 \\ 
% \hline
% 3d-sc & 25.4 & 30.8 & 30.3 & 28.0 & 27.3 & 27.2 & 29.1 & 30.7 & 31.2 \\
% bcc & 21.6 & 20.2 &  20.0 &  21.5 & 21.2 &  21.9 & 23.5 & 24.7 & 25.5 \\
% 4d-sc & 13.8 & 13.0 & 11.2 & 11.1 & 12.2 & 13.6 & 13.5 & 14.6 & 15.1 
% \end{tabularx}
% \label{tab:percentage_reduction}
% \caption{ Reduction of the DMFT N\'eel temperature by lD$\Gamma$A (in percent) at different values of $U$ for the three bipartite lattices under consideration.}
% \end{table}
{\sl fcc lattice:} For the fcc lattice we could find an antiferromagnetic phase transition within DMFT only for $U\!>\!3.0$ due to the strong geometric frustration in this lattice (see inset in Fig.~\ref{fig:phase_diagram}).
This $T_N$ of DMFT is reduced to zero by the nonlocal correlations captured by lD$\Gamma$A for all interaction strengths considered in this study.
In fact, in this lattice a magnetic order at finite temperature as predicted by DMFT is a pure correlation effect as no nesting vector exists in the noninteracting dispersion which would lead to a weak-coupling Slater type magnetism\cite{Ulmke1998,Balla2020,Schick2022,Oitmaa2023}.

\subsection{Weak coupling}
\label{sec:weakcoupling}

\begin{figure}
    \centering
    \includegraphics[width=\linewidth]{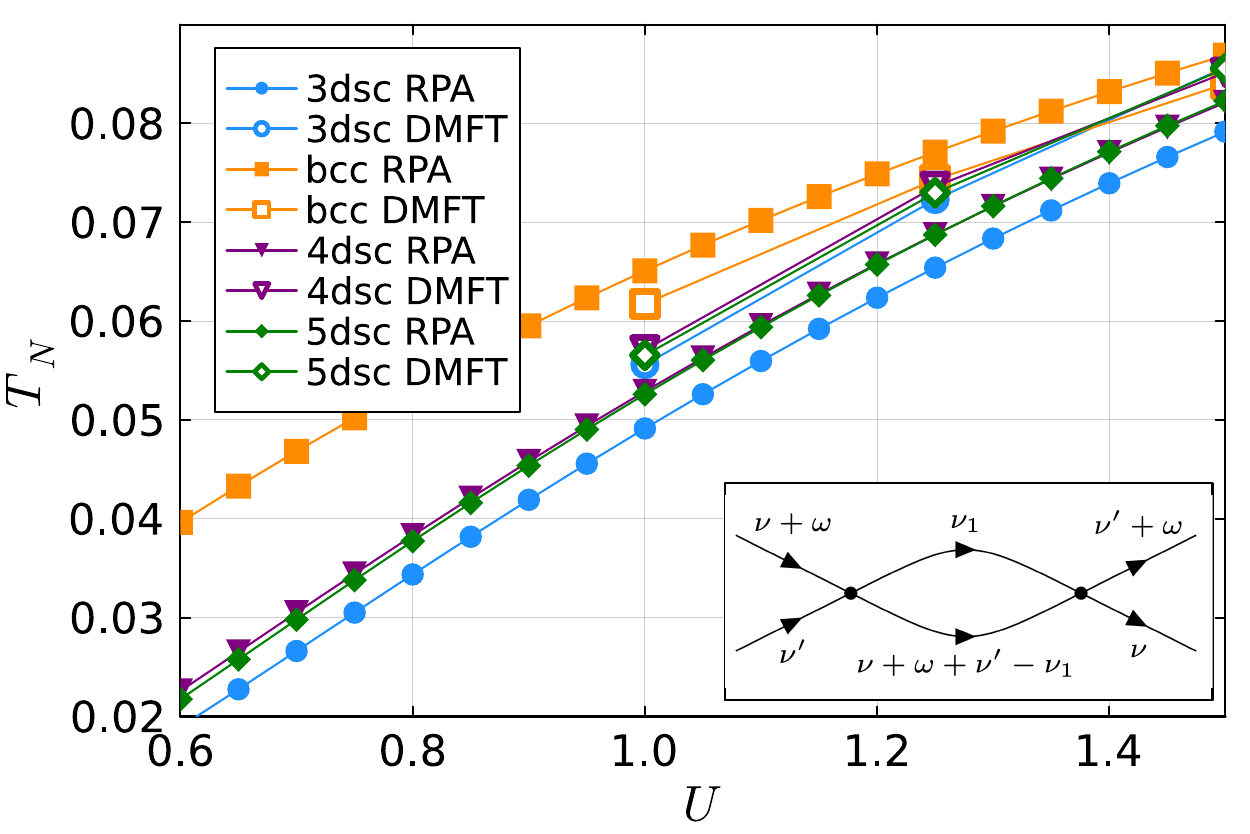}
    \caption{RPA results for $T_N$ for the 3d-sc (filled blue circles), bcc (filled orange squares), 4d-sc (filled violet triangles) and 5d-sc (filled green diamonds) lattice, which are corrected by the second-order diagrams shown in the inset [see also Eq.~\eqref{equ:weakcoupling}]. DMFT results (empty symbols) are given as reference at larger coupling strengths.}
    \label{fig:low_coupling}
\end{figure}

Due to our intentional choice of the hopping parameters  discussed in the previous Secs.~\ref{sec:model}, the transition temperatures of DMFT in Fig.~\ref{fig:phase_diagram}  are very similar for the four bipartite lattices, with an increasing matching in the strong-coupling limit.
For intermediate and especially weak coupling on the other hand small but clearly visible differences can be observed in Fig.~\ref{fig:phase_diagram}, in particular between $T_N$ for the 3d-sc (empty blue circles) and the bcc (empty orange squares) lattices.
More specifically, at the lowest coupling strength $U=1.0$ the DMFT transition temperature for the bcc lattice is larger than the one for the 3d-sc lattice.
On the contrary, at $U=2.0$ the situation is reversed with a slightly larger $T_N$ for the 3d-sc with respect to the bcc lattice.
%\gr{
To verify our DMFT results at weak coupling, where numerical calculations are indeed difficult due to the low values of $T_N$, we have carried out a semi-analytic weak coupling expansion as proposed in Refs.~\cite{vanDongen2002} and \cite{jarrell1997} which should be consistent with the DMFT results for $T_N$ up to $U\sim1.0$ .
%}
Within this approach the simple random phase approximation, where the irreducible vertex in the magnetic channel is replaced by the bare interaction $\Gamma_m^{\nu\nu'\omega}=-U$, is extended to second order in $U$ by considering
% To investigate the weak coupling behavior of every lattice type we perform a second order RPA analysis \cite{Jarrell1997,vanDongen2002}, which is usually in good agreement with DMFT results at small $U$. We are especially interested in the critical temperature of the bcc lattice. As can be seen in Fig.~\ref{fig:phase_diagram}, all DMFT phase transitions are very close (as a result of the units chosen). However, in the low coupling regime, the bcc transition is enhanced in contrast to the other lattice types. \\
% To further investigate this phenomenon, we check whether this is consistent with the second order RPA analysis (Fig. ~\ref{fig:low_coupling}). This method applies a factor to the (commonly used) first order RPA results ($\Gamma_\mathrm{loc} = U$), which is the result of considering the second order order in the interaction for the locale single-channel irreducible vertex:
\begin{equation}
    \Gamma_\mathrm{loc}^{\nu\nu'(\omega=0)} = -U + U^2 \chi_{0,\mathrm{loc}}^{pp,\nu+\nu'},
\label{equ:weakcoupling}
\end{equation}
where $\chi_{0,\mathrm{loc}}^{pp,\nu+\nu'}=\sum_{\nu_1\mathbf{k}_1\mathbf{q}}G_0(\nu_1,\mathbf{k})G_0(\nu_1+\nu+\nu',\mathbf{k}+\mathbf{q})$ is the bare local pairing susceptibility (particle-particle bubble) and $G_0(\nu,\mathbf{k})=1/(i\nu+\mu-\varepsilon_\mathbf{k})$ the noninteracting Green's function.
At weak-to-intermediate coupling this approximation yields results for $T_N$ as a function of $U$ which are in quite good qualitative agreement with our numerical DMFT calculations.
This is demonstrated in Fig.~\ref{fig:low_coupling} where the second-order results for $T_N$ obtained from Eq.~\eqref{equ:weakcoupling} for the 3d-sc (filled blue circles), bcc (filled orange squares), 4d-sc (filled violet triangles) and 5d-sc (filled green diamonds) lattices are reasonably close to the corresponding DMFT data (empty symbols).

More importantly, the approximate calculations also reproduce the correct hierarchy of $T_N$ at weak coupling between the 3d-sc and the bcc lattice as predicted by DMFT.
The results of this perturbative approach are mainly determined by the bare (bubble) magnetic susceptibility which is proportional to the noninteracting density of state $D(\varepsilon)$.
While for the 3d-sc (and also 4d-sc, 5d-sc) $D(\varepsilon)$ is rather featureless, it exhibits a Van Hove singularity at the Fermi level for the bcc lattice. 
Considering the RPA like expression for $T_N\sim e^{-\frac{1}{UD(0)}}$, which holds also for the approximation in Eq.~\eqref{equ:weakcoupling}, a higher $D(0)$ as in the bcc lattice implies indeed a higher N\'eel temperature.
At the same time, this higher density of states provides more phase space for electron-electron scattering which increases the correlation effects at larger values of $U$.
These enhanced correlations tend to suppress $T_N$\cite{DelRe2021} for the bcc lattice at intermediate coupling strength $U\sim2.0$ stronger than in the 3d-sc lattice, where such correlation effects are weaker due to the absence of a van Hove singularity.
This explains the higher values of $T_N$ for the 3d-sc with respect to the bcc lattice in this parameter regime.

Let us finally mention that the hierarchy of curves in the phase diagram obtained from lD$\Gamma$A at weak coupling, in particular the equal values of $T_N$ for the bcc and 4d-sc lattices at $U=1.0$, cannot easily be explained.
In this respect we should mention, that the low coupling values for $T_N$ provided by lD$\Gamma$A with a $\lambda$ correction only in the spin channel might be underestimated as a renormalization of charge fluctuations is neglected.
More accurate results require an improved treatment of charge fluctuations as it has been recently developed for lD$\Gamma$A in Refs.~\cite{Stobbe2022,Titvinidze2025}.

% As expected, the effect is also apparent in RPA and even more pronounced. \\
% We assume this to be the result of the van Hove singularity present in the bcc lattice. This leads to a large density of states at the Fermi surface and thus, a large $\chi_0$, resulting in an enhancement of the critical temperature. This is less drastic in DMFT since the inclusion of correlation effects broadens the van Hove singularity. 

%\gr{This section will be filled according to Marvin's analytical/numerical results for weak coupling.}

\subsection{Violation of sum rules}
\label{sec:sumrules}

The ladder D$\Gamma$A includes nonlocal correlation effects in an effective way by restoring the sum rule in Eq.~\eqref{eq:pauli_principle}, which is violated when using DMFT in finite dimensions.
Hence, the magnitude of this violation $d_\text{PP}$, which is just the difference between the right- and left-hand side of Eq.\eqref{eq:pauli_principle}, can be used as an indicator for the reliability of DMFT results for different lattices and parameter regimes
\begin{equation}
\label{equ:dpp}
d_\text{PP}=\frac{1}{2}\sum_{\omega\mathbf{q}}\left(\chi_{d,\mathbf{q}}^\omega + \chi_{m,\mathbf{q}}^\omega\right)-\frac{n}{2}(1-\frac{n}{2})
\end{equation}
We have hence calculated $d_\text{PP}$ for different coupling strengths as a function of $T$ to analyze how the violation of this sum rule depends on the lattice type and the proximity to the antiferromagnetic phase transition.
Note that for $T\rightarrow\infty$ this quantity should approach zero as DMFT becomes exact in this limit.

\begin{figure}[t!]
    \includegraphics[width = 0.49\textwidth]{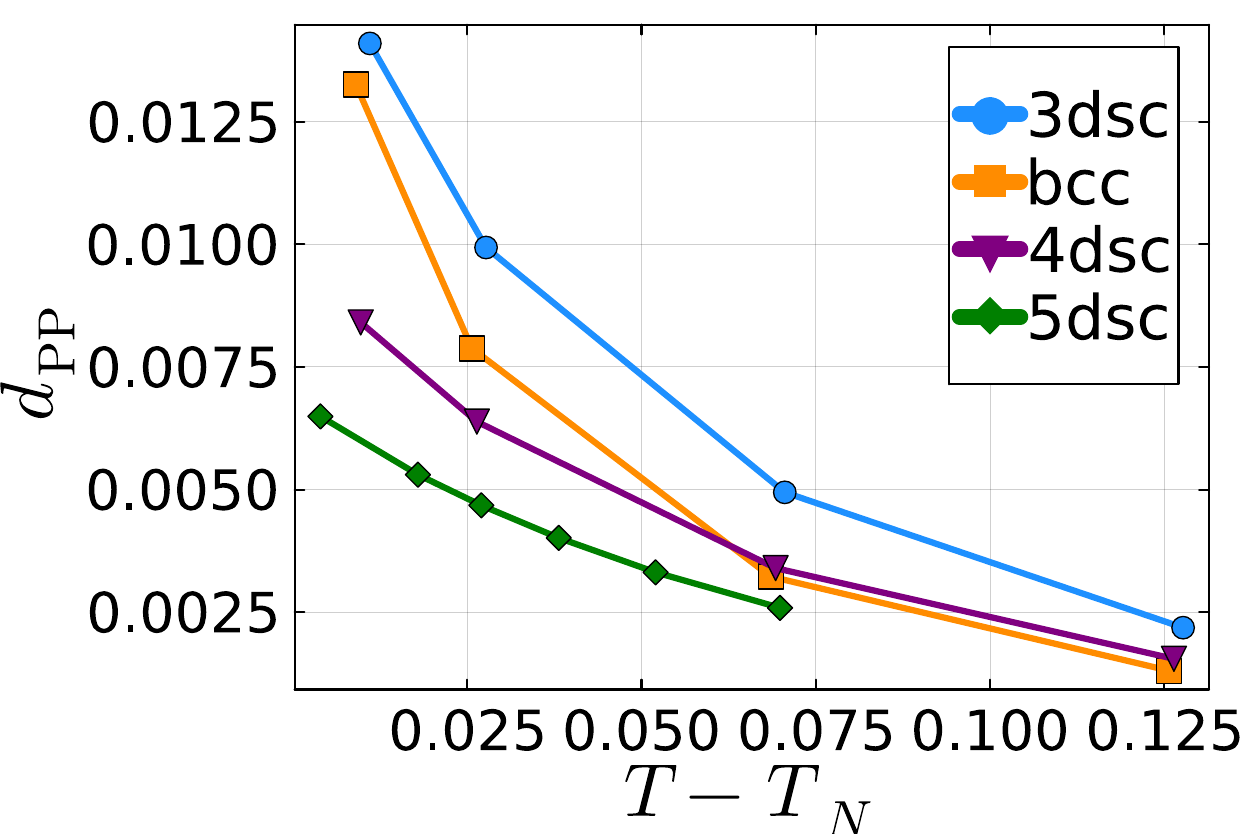}
    \includegraphics[width = 0.49\textwidth]{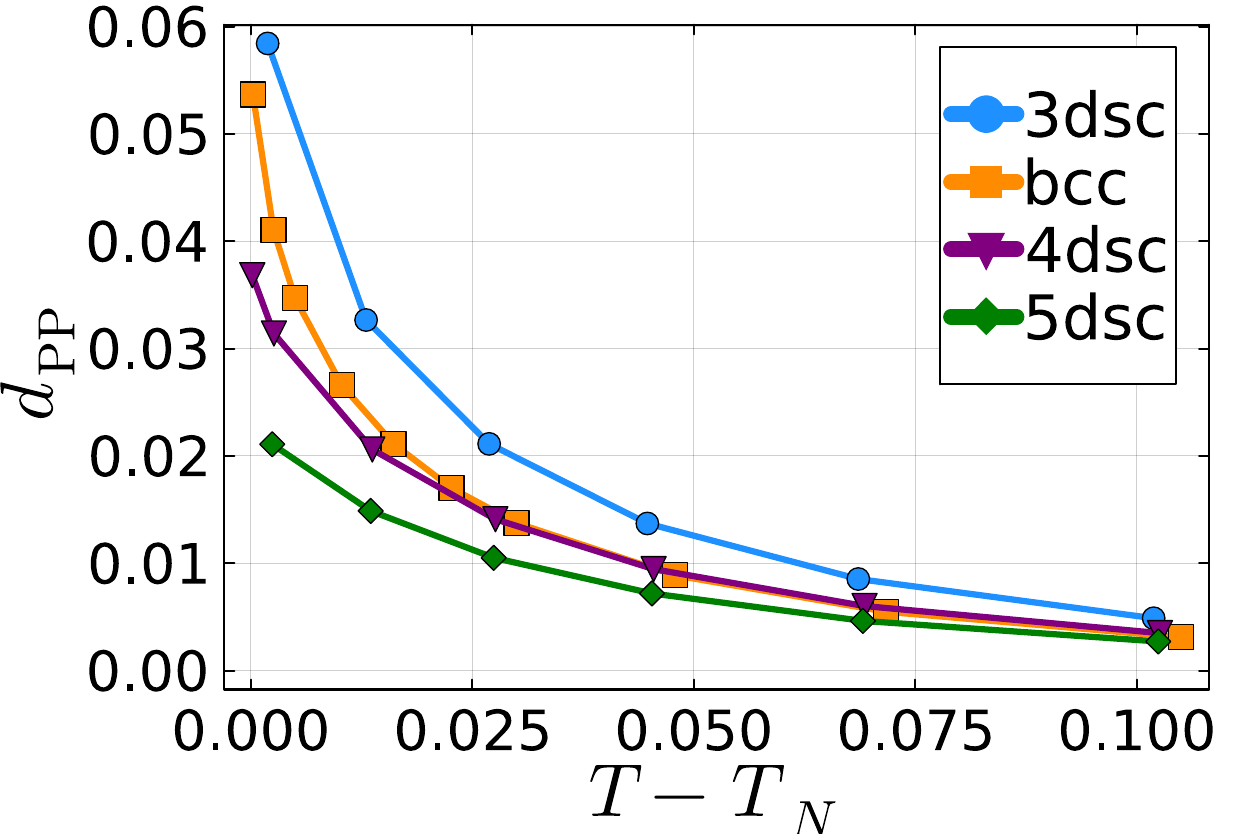}
    \caption{
    %\gs{to which values should $d_PP$ go at large $T$? How can one quantify the violation? Maybe write in the caption and in the text how large should $d_PP$ be.} 
    Violation of the Pauli principle in DMFT [see Eq.~\eqref{equ:dpp}] on the 3d-sc (blue circles), bcc (orange squares), 4d-sc (violet triangles) and 5d-sc (green diamonds) lattices at $U= 1.25$ (left panel) and  $2.0$ (right panel). For $T\rightarrow\infty$ $d_\text{PP}(T)$ approaches zero as DMFT becomes exact in this limit.
    %As expected DMFT gets less accurate the closer we get to the phase transition for both lattice. Even thought they have the same coordination number ($Z=8$), DMFT results are still better for the true four dimensional lattice.
    }
    \label{fig:pp_violation}
\end{figure}

{\sl Bipartite lattices:} In Fig.~\ref{fig:pp_violation} we plot the violation $d_\text{PP}$ of the Pauli principle in Eq.~\eqref{equ:dpp} for all bipartite lattices for $U\!=\!1.25$ (left panel) and $U\!=\!2.0$ (right panel), respectively, as a function of $T\!-\!T_N$ where $T_N$ denotes the transition temperature of the respective lattice.
We clearly observe a pronounced increase of the violation upon approaching the transition temperature of DMFT. 
This is indeed the expected behavior as the increasingly strong nonlocal fluctuations in the vicinity of the phase transition are not adequately captured in DMFT. 
The hierarchy between the four bipartite lattices follows their different coordination numbers and spatial dimensions.
The largest violation is observed for the 3d-sc lattice with a coordination number of $6$ followed by the bcc lattice with a coordination number of $8$.
Though it has the same coordination number as the bcc lattice, substantially lower increase of $d_\text{PP}$ in the vicinity of $T_N$  is observed for the tesseract case, due to the additional dimension. Finally, the 5d-sc lattice shows the smallest violation of the Pauli principle at coordination number 10. The evolution of $d_\text{pp}$ convincingly illustrates the systematically improved quality of DMFT results for higher coordination number and spatial dimensions.

\section{Conclusions and Outlook}
\label{sec:conclusions}

In this paper we have investigated the impact of nonlocal correlations in strongly correlated electron systems for different lattice geometries.
To this end we have considered the Hubbard model for the 3d-sc, bcc, 4d-sc, 5d-sc, and fcc lattices.
The first four lattices are bipartite and, hence, feature a strong tendency to antiferromagnetic order, while in the fcc lattice magnetic fluctuations are strongly suppressed due to geometric frustration.
In a first step, we have analyzed the transition temperatures $T_N$ to the magnetically ordered state exploiting DMFT, which takes into account local correlations only.
In a second step, we have demonstrated how these transitions temperatures are reduced by the inclusion of nonlocal correlation effects by means of lD$\Gamma$A.
We found a clear hierarchy for the the differences between DMFT and lD$\Gamma$A which decrease with increasing coordination number and dimension of the lattice.
This is indeed the expected behavior as DMFT becomes exact in the limit of infinite coordination number or dimension.
To complete our analysis of $T_N$ for low values of $U$ we have used a low-coupling expansion of $T_N$ which has allowed us to understand the enhanced value of the transition temperature in the bcc lattice in terms of the van Hove singularity emerging at the Fermi level in this system.

We have also investigated the critical behavior close to $T_N$ and, in particular, calculated the critical exponent $\gamma$ for all lattice types.
While for the 4d-sc and 5d-sc lattices we obtain $\gamma\sim 1$ (except for logarithmic corrections), which is consistent with scaling theory in dimensions $d\ge 4$, our numerical data indicate, (within the specific framework of Moriya corrected lD$\Gamma$A) definitely larger values of critical exponents of the three-dimensional bipartite lattices, numerically compatible with the Berlin-Kac model universality class.

Finally, we have quantified the violation of a specific sum rule for the magnetic susceptibilities within DMFT, further supporting the observed hierarchy for the impact of nonlocal correlations for the different lattice geometries according to their coordination number/dimension.

 Our work can serve as a useful ``compass'' for the quantitative description of emerging magnetic orders of correlated electrons in different lattice geometries for both experimental and theoretical studies.
Our analysis can be further improved by more sophisticated description of nonlocal correlations which takes into account also charge fluctuations in the weak-to-intermediate coupling regime, which has been recently proposed in Refs.~\cite{Stobbe2022} and \cite{Titvinidze2025}.
Moreover, the study of the fcc lattice could be extended to other fillings where a wide variety of different phases in observed by weak-coupling methods.

%\bigskip
%\noindent
%Check ``Reviewing'' Modus ;-) we need an Acknowledgment section

{\sl Acknowledgements}---We thank S. Andergassen, B. Sbierski, and L. Del Re for useful discussions. M.L. and G.R. acknowledges financial support from the Deutsche Forschungsgemeinschaft (DFG) through Projects No. 407372336 and Project P8 of the FOR 5249 [QUAST] No. 449872909.  A.T. acknowledges financial support from the Austrian Science Fund FWF, through the projects with Grant-DOIs: 10.55776/I5868 and 10.55776/KIN2563725,corresponding to P01 of the QUAST Research Unit of DFG (for 5249). 
G.S. acknowledges financial support by the DFG through Project No.~468199700 as well as through Project P5 of the FOR 5249 [QUAST] No.~449872909.
The authors gratefully acknowledge the computing time granted by the Resource Allocation Board and provided on the supercomputers Lise and Emmy/Grete at NHR@ZIB and NHR@G\"ottingen as part of the NHR infrastructure. These centers are jointly supported by the Federal Ministry of Education and Research and the state governments participating in the NHR (www.nhr-verein.de).
The calculations for this research were conducted with computing resources under the Project hhp00048.

\begin{figure}[t!]
    \includegraphics[width = 0.49\linewidth]{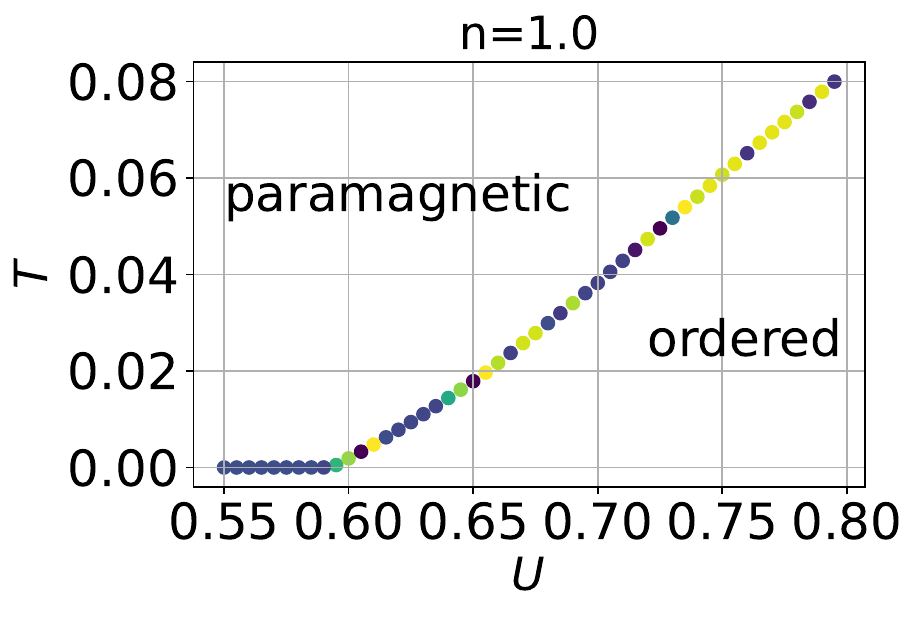}
    \includegraphics[width = 0.49\linewidth]{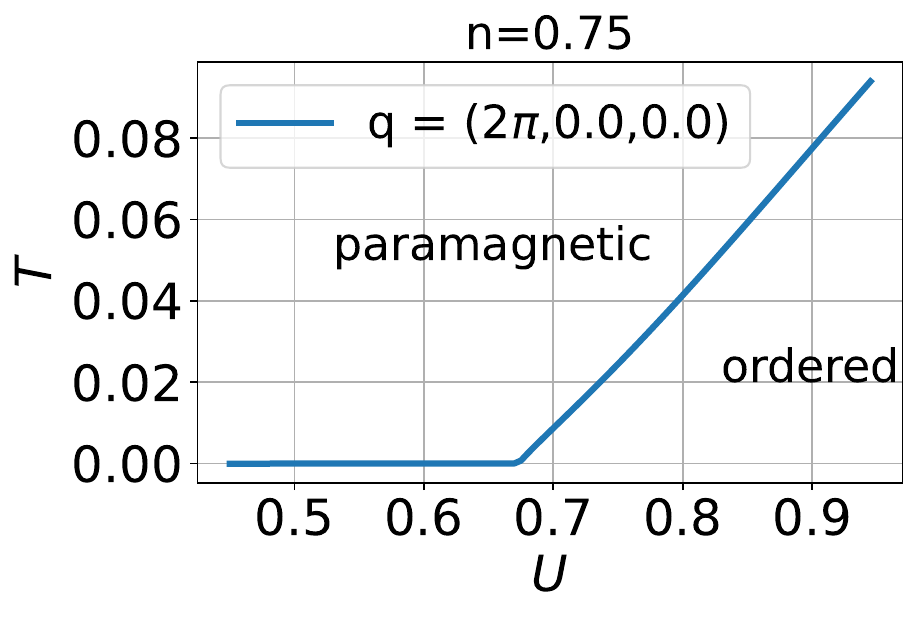}
    \caption{%\gs{numbers and symbols way too small!!!} 
    RPA $U$ vs $T$ phase diagram of the fcc lattice at $n=1.0$ (left panel) and $n=0.75$ (right panel). In the left panel, different colors of the transition points indicate different ordering vectors.
    %we only show some sample ordering vectors, however almost every data point in $U$ orders at least slightly differently.}
    }\label{fig:RPA_fcc}
\end{figure}

\appendix
\vspace*{0.6cm}

\section{RPA results for the fcc lattice}
\label{sec:rpa}

The fcc lattice features a strong geometric frustration which leads, in general, to a substantial suppression of magnetically ordered phases. Moreover, the absence of a nesting vector prevents a clearly dominating type of magnetic order but, instead, facilitates a large number of different ordering structures such as antiferromagnetic, stripe or spiral patterns depending on filling $n$, interaction strength $U$ and temperature $T$.
According to RPA calculations for the ground state ($T=0$) phase diagram in Ref.~\cite{Igoshev2015} these phases are separated by either sharp second-order phase transitions or regions of phase separation.
Since some of the regions with a well-defined ordered ground state are rather narrow in the $U$ vs $n$ phase diagram we have carried out RPA calculations at finite temperature to identify the fillings where a stable ordered phase is expected to exist also at $T>0$.

%When investigating the fcc lattice one of the struggles we faced was the rich phase diagram. As a result of its geometric frustration, there is no clear antiferromagnetic ordering, leading to a variety of different spin orders including $\mathbf{q} = (0,0,2\pi)$, $\mathbf{q} = (\pi,\pi,\pi)$ and $\mathbf{q} = (0,\pi,2\pi)$, as well as a variety of different spiral phases. Additionally, in between these phases one finds a lot of crossover regions. \\
%To find a stable phase we extend the ground state Hartree-Fock phase diagram found by \cite{Igoshev2015} to finite temperature, only for some dopings. Since the purpose of this analysis is only to find a stable area of the phase diagram for the DMFT calculations, we use RPA, neglecting most of the correlation effects suppressing the Neel temperature in DMFT and D$\Gamma$A. Additionally, this allows us to compare the results very closely with \cite{Igoshev2015}. \\ 

In a first step, we have investigated $n=1.0$ for which the $U$ vs. $T$ phase diagram is shown in the left panel of Fig. \ref{fig:RPA_fcc}. 
The filled circles indicate the RPA transition temperature to an magnetically ordered state whereas the color corresponds to the ordering vector $\mathbf{q}_N$.
While at low values of $U$ no order emerges due to the strong geometrical frustration, we observe a finite transition temperature $T$ for $U\gtrsim 0.6$.
As the different colors of the transition points indicate, the ordering vector $\mathbf{q}_N$ changes for virtually any $U$ which is incompatible with a well-defined ordered state.
This is consistent with the ground state analysis from Ref.~\cite{Igoshev2015} where $n=1.0$ marks a transition line between various ordered, incommensurate and phase separated states in the $U$ vs $n$ phase diagram.

We have, hence, carried out calculations also for $n=0.75$ where the mean-field results of Ref.~\cite{Igoshev2015} predict a clear-cut ordered state at finite $U$ with the ordering vector $\mathbf{q}_N=(0,0,2\pi)$ at $T=0$.
Our RPA data in the right panel of Fig.~\ref{fig:RPA_fcc} confirm that such a stable magnetic phase survives also at finite temperatures for interaction values $U\gtrsim 0.65$  which is in reasonable good agreement with the before mentioned ground state results.
Consequently, we have carried out our DMFT and lD$\Gamma$A calculations for this filling in our manuscript.

%Here we can clearly see a number of different ordering vectors at different interaction strengths. This is consistent with the results of \cite{Igoshev2015} that predict a coexistence region between at least three different ordering vectors at $n=1.0$, with some phases having a spiral, leading to an even wider variety of ordering vectors. This instability makes DMFT and D$\Gamma$A calculations a lot more complicated. Thus we consider a doped system, in this case $n=0.75$, which should firmly lie in the phase $\mathbf{q} = (0,0,2\pi)$. We confirm this with our RPA calculations (see the right panel of Fig. \ref{fig:RPA_fcc}). As a result, we perform all of our DMFT and D$\Gamma$A calculations at $25\%$ hole doping.\\

\FloatBarrier

%apsrev4-2.bst 2019-01-14 (MD) hand-edited version of apsrev4-1.bst
%Control: key (0)
%Control: author (8) initials jnrlst
%Control: editor formatted (1) identically to author
%Control: production of article title (0) allowed
%Control: page (0) single
%Control: year (1) truncated
%Control: production of eprint (0) enabled
%

%\bibliographystyle{apsrev4-2}
%\bibliography{georg_thesis} 

\end{document}